\documentclass[sigconf]{acmart}

\AtBeginDocument{%
  \providecommand\BibTeX{{%
    \normalfont B\kern-0.5em{\scshape i\kern-0.25em b}\kern-0.8em\TeX}}}

\copyrightyear{2023}
\acmYear{2023}
\setcopyright{rightsretained}
\acmConference[KDD '23] {Proceedings of the 29th ACM SIGKDD Conference on Knowledge Discovery and Data Mining}{August 6--10, 2023}{Long Beach, CA, USA.}
\acmBooktitle{Proceedings of the 29th ACM SIGKDD Conference on Knowledge Discovery and Data Mining (KDD '23), August 6--10, 2023, Long Beach, CA, USA}
\acmPrice{}
\acmISBN{979-8-4007-0103-0/23/08}
\acmDOI{10.1145/3580305.3599509}

\title{The Name of the Title is Hope}

\usepackage{algorithmic}
\usepackage{algorithm}
\usepackage{natbib}
\usepackage{graphicx}
\usepackage{subcaption}
\usepackage{booktabs} 
\usepackage{amsfonts,graphics,epsfig,verbatim,bm}
\usepackage{latexsym,amsmath,url,amsbsy}
\usepackage{graphicx}
\usepackage{booktabs,setspace}
\usepackage{caption}
\usepackage{multirow}
\usepackage[utf8]{inputenc} 
\usepackage[T1]{fontenc}    
\usepackage{hyperref}       
\usepackage{url}            
\usepackage{booktabs}       
\usepackage{amsfonts}       
\usepackage{nicefrac}       
\usepackage{microtype}      
\usepackage{xcolor}         
\usepackage{adjustbox}
\usepackage{pdfpages}
\usepackage{hyperref}

\usepackage{amsmath}
\usepackage{mathtools}
\usepackage{amsthm}

\theoremstyle{plain}
\newtheorem{theorem}{Theorem}[section]
\newtheorem{proposition}[theorem]{Proposition}
\newtheorem{lemma}[theorem]{Lemma}

\theoremstyle{definition}

\theoremstyle{remark}
\newtheorem{remark}[theorem]{Remark}

\begin{document}

\title[Markov Random Field Mixture of Finite Mixtures]{Spatial Clustering Regression of Count Value Data via Bayesian Mixture of Finite Mixtures}

\author{Peng Zhao}
\email{pzhao@tamu.edu}
\orcid{0000-0001-7043-0389}
\affiliation{%
  \institution{Texas A\&M University}
  \state{Texas}
  \country{USA}
}

\author{Hou-Cheng Yang}
\email{houchengyang4343@gmail.com}
\orcid{0000-0002-8679-4280}
\affiliation{%
  \institution{Florida State University}
  \state{Florida}
  \country{USA}}

\author{Dipak K Dey}
\email{dipak.dey@uconn.edu}
\affiliation{%
  \institution{University of Connecticut}
  \state{Connecticut}
  \country{USA}
}

\author{Guanyu Hu}
\email{nealguanyu@gmail.com}
\orcid{0000-0003-1410-1665}
\affiliation{%
 \institution{Department of Biostatistics and Data Science and Center for Spatial-Temporal Modeling for Applications in Population Sciences (CSMAPS), The University of Texas Health Science Center at Houston}
 \state{Texas}
 \country{USA}}


\begin{abstract}
Investigating relationships between response variables and covariates in areas such as environmental science, geoscience, and public health is an important endeavor. Based on a Bayesian mixture of finite mixtures model, we present a novel spatially clustered coefficients regression model for count value data. The proposed method detects the spatial homogeneity of the Poisson regression coefficients. A Markov random field constrained mixture of finite mixtures prior provides a regularized estimator of the number of clusters of regression coefficients with geographical neighborhood information. As a by-product, we also provide the theoretical properties of our proposed method when the Markov random field is exchangeable. An efficient Markov chain Monte Carlo algorithm is developed by using the multivariate log gamma distribution as a base distribution. Simulation studies are carried out to examine the empirical performance of the proposed method. Additionally, we analyze Georgia's premature death data as an illustration of the effectiveness of our approach. The supplementary materials are provided on GitHub at \url{https://github.com/pengzhaostat/MLG_MFM}.
\end{abstract}




\maketitle

\section{Introduction}
Spatial regression models have been widely used in various fields, such as environmental science \citep{hu2018stat, yang2019bayesian, yang2021bayesian}, biological science \citep{zhang2011bayesian}, and econometrics \citep{brunsdon1996geographically, yang2022spatial}, to explore the relationship between a response variable and a set of predictors over a region. One of the key objectives of spatial regression models is to capture the spatial dependence structure of the response variable. Spatial random effects are typically incorporated through intercepts, while the regression coefficients are assumed to be constant over space in both linear models \citep{cressie1992statistics} and generalized linear models \citep{diggle1998model}. To capture spatially varying patterns in regression coefficients, \cite{brunsdon1996geographically} proposed geographically weighted regression (GWR). This idea has been further extended in subsequent works by \cite{hu2018modified, ma2019geographical, xue2019geographically}. Additionally, \cite{gelfand2003spatial} incorporated spatial Gaussian processes into linear regressions to construct a spatially varying coefficients regression model. However, these approaches assume that each location has its own set of regression parameters, which can sometimes lead to overfitting. The detection of clustered covariate effects has significant benefits in various fields, including environmental science, spatial econometrics, and disease mapping. For example, different regions of a country may exhibit distinct economic conditions and development patterns. From a modeling perspective, grouping more advanced regions and less developed regions into separate clusters can lead to a more parsimonious model.

\subsection{Related Work and Challenges}
Spatial cluster detection methods, such as the scan statistic-based method \citep{kulldorff1995spatial, jung2007spatial}, provide a solution for detecting spatial heterogeneity. Another important approach for spatial heterogeneity detection is to utilize the Bayesian framework to identify spatial clusters \citep{carlin2014hierarchical, li2010nonparametric}. These two approaches primarily focus on estimating cluster configurations of spatial responses. Recently, methods for detecting clusters of spatial regression coefficients have been proposed to assess the homogeneity of covariate effects among sub-areas \citep{lee2017cluster, leespatial2019} using spatial scan statistics. From a graph theory perspective, \cite{li2019spatial} incorporated spatial neighborhood information based on minimum spanning trees into a penalized approach to identify spatially clustered coefficients. However, the existing literature mainly focuses on Gaussian data within the linear model framework. In many social and environmental applications, Poisson regression for count responses plays a crucial role \citep{bradley2018computationally}.

Developing clustering algorithms for regression coefficients under Poisson models presents several significant challenges. First, specific spatial contiguity constraints need to be imposed on the clustering configuration to facilitate interpretations in the spatially clustered coefficients regression. Additionally, in many regional science applications, spatially contiguous constraints should not dominate the overall clustering configuration. In other words, the clustering results should include both spatially contiguous and spatially disconnected patterns. While the aforementioned methods \citep{lee2017cluster, leespatial2019, li2019spatial} guarantee spatial contiguity, they fail to obtain globally discontiguous clusters that allow two clusters with long geographical distances to be considered part of the same cluster. Furthermore, \cite{anderson2017spatial} discusses Poisson regression with spatially clustered intercepts and slopes but does not impose a spatial contiguity constraint.

Another crucial consideration in clustering algorithms is the estimation of the number of clusters. Bayesian inference provides a probabilistic framework for simultaneous inference of the number of clusters and the clustering configurations. Nonparametric Bayesian approaches, such as the Dirichlet process mixture (DPM) model \citep{ferguson1973bayesian}, offer choices to estimate the number of clusters and the clustering configurations simultaneously. However, the Dirichlet process mixture model suffers from inconsistency issues \citep{miller2013simple}, which leads to challenges in obtaining a consistent estimator of the number of clusters. To address the problem of over-clustering in the DPM, several approaches in the literature \citep{miller2018mixture, xie2019bayesian, lu2018reducing} propose different ideas to obtain consistent estimators of the number of clusters. However, these existing methods do not utilize spatial information, such as neighborhood relationships, despite the potential to enhance clustering performance. According to Tobler's first law of geography \citep{tobler1970computer}, "Everything is related to everything else, but near things are more related than distant things." Considering a similar pattern in the data due to similar environmental circumstances is reasonable. Although it may be challenging to incorporate arbitrary types of spatial information with consistent guarantees on the number of clusters, exploring specific types of dependency structures without compromising consistency is still of interest.

\subsection{Contributions}

To address these challenges, we propose a novel approach called the Markov random field (MRF) constrained Mixture of Finite Mixtures (MFM) model (MRF-MFM) to capture the spatial homogeneity in regression coefficients for Poisson models. Our Bayesian method for spatially clustered coefficients Poisson regression incorporates geographical information using the MRF constrained MFM model. This enables us to capture both locally spatially contiguous and globally discontiguous clusters simultaneously. We develop a Gibbs sampler that facilitates efficient full Bayesian inference on the number of clusters, mixture probabilities, and other modeling parameters for Poisson regression, leveraging the multivariate log gamma (MLG) process \citep{bradley2018computationally}. Through simulations and an analysis of premature deaths data in the state of Georgia, we demonstrate the excellent numerical performance of our proposed mixture models. Additionally, we establish a consistency result for the posterior estimates of the cluster number and associated modeling parameters when the Markov random field is assumed to be exchangeable.

Our proposed method has several unique aspects. First, the introduction of the Markov random field into the mixture of finite mixtures model for spatial cluster coefficients regression is a novel idea. This approach has wide applicability in spatial statistics applications, such as environmental science and geographical analysis, providing a valuable alternative to existing literature that primarily relies on penalized regression or scan statistics. We provide a detailed comparison with related literature (e.g., \cite{li2019spatial, anderson2017spatial}) in Section~\ref{sec:comparison} of the supplement. Second, by adopting a full Bayesian framework, our clustering results offer useful probabilistic interpretations. Moreover, our developed posterior sampling scheme ensures efficient computation. Third, our theoretical result is among the first of its kind for mixture models under the exchangeable assumption. The posterior consistency result not only justifies the excellent empirical performance (e.g., regularization on the number of clusters) but also connects with existing theoretical findings on mixture models in general.

\section{Methodology}\label{sec:method}

\subsection{Clustered Poisson Regression and Mixture of Finite Mixtures}
Consider a Poisson regression model with spatially varying coefficients as
follows
\begin{equation}
	y(\bm{s}_i)\sim \text{Poisson}(\exp(\bm{X}(\bm{s}_i)\bm{\beta}_{z_i})),~~
	i=1,\cdots,n,
\end{equation}
where $\bm{X}(\bm{s}_i)$ is a $n \times p$ covariates matrix, $z_i \in \{1,\cdots,k\}$ are labels of clusters, $\bm{\beta}_{z_i} = \bm{\beta}(s_i)$ is a~$p$ dimensional regression coefficients at
location $\bm{s}_i$. From \cite{gelfand2003spatial}, a Gaussian process prior
can be assigned on regression coefficients to obtain spatially varying pattern.
Compared with spatially varying pattern, heterogeneity pattern of covariate
effects over subareas is also universally discussed in many different fields,
such as real estate applications, spatial econometrics, and environmental
science. 

In the popular Chinese restaurant process, $z_i$, $i=2, \ldots, n$ are
defined through the following conditional distribution \cite{ferguson1973bayesian}:
\begin{align}\label{eq:crp}
\begin{aligned}
&	P(z_{i} = c \mid z_{1}, \ldots, z_{i-1})  \\
	& \propto   
	\begin{cases}
		\left|c\right|  , \quad \text{at an existing table labeled}\, c\\
		\gamma, \quad \text{if} \, $c$\,\text{is a new table}
	\end{cases},
	\end{aligned}
\end{align}
where $\left|c\right|$ is the size of cluster $c$, and $\gamma$ is a concentration parameter of Dirichlet Process.

While CRP has a very attractive feature of simultaneous estimation on the number
of clusters and the cluster configuration, a striking consequence of this has
been recently discovered \cite{miller2013simple} where it is shown that the
CRP produces extraneous clusters in the posterior, leading to inconsistent
estimation of the {\em number of clusters} even when the sample size grows to
infinity. 
A modification of the CRP called Mixture of finite mixtures (MFM)
model is proposed to circumvent this issue \cite{miller2018mixture}:
\begin{align}\label{eq:MFM}
\begin{aligned}
	k \sim p(\cdot), \quad (\pi_1, \ldots, \pi_k) \mid k \sim \mbox{Dir} (\gamma,
	\ldots, \gamma), \\
	\quad z_i \mid k, \pi \sim \sum_{h=1}^k  \pi_h \delta_h,\quad
	i=1, \ldots, n, 
	\end{aligned}
\end{align}
where~$p(\cdot)$ is a proper probability mass function on~$\{1, 2, \ldots, \}$, $\gamma$ is a concentration parameter of Dirichlet Process
and~$\delta_h$ is a point-mass at~$h$. Compared to the CRP, the introduction of new tables is slowed down by the factor~$V_n(t+1)/ V_n(t)$, which allows a model-based pruning of the tiny extraneous clusters. 

The coefficient~$V_n(t)$ is precomputed as:
\begin{align}\label{vt} 
	\begin{split}
		V_n(t) &= \sum_{k=1}^{+\infty}\dfrac{k_{(t)}}{(\gamma k)^{(n)}} p(k),
	\end{split}         
\end{align} 
where $k_{(t)}=k(k-1)\ldots(k-t+1)$, and $(\gamma k)^{(n)} = {\gamma k}(\gamma
k+1)\ldots(\gamma k+n-1)$. $z_i,
i=2, \ldots, n$ under \eqref{eq:MFM} can be defined in a P\'{o}lya urn scheme
similar to CRP:
\begin{align}\label{eq:mcrp}
\begin{aligned}
&	P(z_{i} = c \mid z_{1}, \ldots, z_{i-1}) \\
	&\propto 
	\begin{cases}
		\left|c\right| + \gamma  , \quad \text{at an existing table labeled}\, c.\\
		\gamma V_n(t+1)/ V_n(t), \quad \text{if} \, c\,\text{is a new table}.
	\end{cases},
	\end{aligned}
\end{align}
where $t$ is the number of existing clusters.  

\subsection{Introducing Dependency on the Base Measure}
Recall that the full model for MFM is
\begin{equation}\label{MFM}
	\begin{split} 
		& K \sim p_{K}, \,\,\mbox{where} \,\, p_{K}\,\, \mbox{is a p.m.f. on} \{1,2,
		\ldots\} \\
		& \left(\pi_{1}, \ldots, \pi_{k}\right) \sim
		\operatorname{Dirichlet}_{k}(\gamma, \ldots, \gamma)\\
	& z_{1}, \ldots, z_{n} \stackrel{\mathrm{iid}}{\sim} \pi, \,\,\mbox{given}\,\,
		\pi\\
		&\bm\beta_{1}, \ldots, \bm\beta_{k}  \stackrel{\mathrm{iid}}{\sim} H,
		\,\,\mbox{given}\,\, K=k\\
		&y_{j} \sim f_{\beta_{z_{j}}}\,\, \mbox{independently for} \,\, j=1, \ldots,
		n,\,\, 
	\end{split}
\end{equation} 
where $H$ is the base distribution for $\beta$.
The main insight of MFM is introducing a prior  on the length of the
Dirichlet distribution, and thus renders some regularization on the number of
clusters created.  However, the fourth step
in the model, where i.i.d. samples are obtained from a base measure, fails to
incorporate any dependency structure. 

Inspired by
\cite{orbanz2008nonparametric}, we apply the pairwise MRF in the level of
coefficients to bring in interactions. With the assistance of Markov random field modeling, our  MRF-MFM can incorporate more broad types
of base measures. Consider an undirected random
graph $G=(V,E,W)$, where $V=\{v_1,\ldots,v_n\}$ is the vertex set while $E$ is
the set of graph edges, with weights  $W$ on the corresponding edges. Each
vertex $v_i$ is associated with a random variable $\bm{\beta}_i$ for
$i=1,2,\ldots,k$.
The pairwise MRF model is defined as 
\begin{equation}\label{Z_H:def}
	\begin{aligned}
		&\Pi\left(\bm{\beta}_{1}, \ldots, \bm{\beta}_{k}\right)=\exp \left\{\sum_{i \in E}
		H_{i}\left(\bm{\beta}_i\right)+\sum_{(i,j)\in E,j \neq i}  H_{ij} (\bm{\beta}_{i}
		\bm{\beta}_{j}) \right.\\
	&\left.	-A\left(W\right)\right\} 
		=\frac{1}{Z_{H}} \exp \left(H\left(\bm{\beta}_{1}, \ldots,
		\bm{\beta}_{k}\right)\right),
	\end{aligned}
\end{equation}
where $Z_H$ is the normalizing constant. For example, for a Gaussian MRF,
$H_i(\bm{\beta}_i)=-W_{ii}\bm{\beta}_i^2/2$ and $H_{ij} (\bm{\beta}_{i}
\bm{\beta}_{j})=-W_{ij}\bm{\beta}_{i} \bm{\beta}_{j}/2$; while for a binary MRF, i.e.,
the celebrated Ising model,  $H_i(\bm{\beta}_i)=W_{ii}\bm{\beta}_i$ and $H_{ij}
(\bm{\beta}_{i} \bm{\beta}_{j})=W_{ij}\bm{\beta}_{i} \bm{\beta}_{j}$.
We can then decompose the pairwise MRF into a vertex-wise term~$P$ and an
interaction term~$M$, then
\begin{equation}\label{M:def}
	\begin{aligned}
		\Pi\left(\bm{\beta}_{1}, \ldots, \bm{\beta}_{k}\right) &\propto P\left(\bm{\beta}_{1},
		\ldots, \bm{\beta}_{k}\right) M\left(\bm{\beta}_{1}, \ldots, \bm{\beta}_{k}\right),
		\,\,\,\mbox{with} \\
		P\left(\bm{\beta}_{1}, \ldots, \bm{\beta}_{k}\right) & :=\frac{1}{Z_{P}} \exp
		\left\{\sum_{i} H_{i}\left(\bm{\beta}_{i}\right)\right\}; \\
		M\left(\bm{\beta}_{1},
		\ldots, \bm{\beta}_{k}\right) & :=\frac{1}{Z_{M}} \exp \left\{\sum_{C \in
			\mathcal{C}_{2}} H_{C}\left(\bm{\beta}_{C}\right)\right\}, \end{aligned}
\end{equation}
where $\mathcal{C}_{2}:=\{C \in \mathcal{C}\mid  s.t.: |C|=2\}$ and $\mathcal{C}$
is the set of all cliques for the random graph $(V,E,W)$. For the spatial
clustered coefficient regression, we study
the component $P$ defined in equation~\eqref{M:def} with a MFM prior.  Our next theorem provides the generalized urn-model induced by MRF-MFM, thus a
collapsed Gibbs sampler can be applied.

%
%

\begin{theorem}
	\label{thm:urn}
	Suppose the data generating process follows equation~\eqref{MFM} with $H$ replaced by the Markov random field  $\Pi(\bm\beta_1,...,\bm\beta_k)$ in equation~\eqref{M:def}. If $P$ is
	a continuous distribution and $n_0>1$, the distributions of $\bm{\beta}_{n_0}$ given
	$\bm{\beta}_1,\ldots,\bm{\beta}_{n_0-1}$ is proportional to
	\begin{equation}
		\frac{V_{n_0}(t+1) \gamma}{V_{n_0}(t) }
		P(\bm{\beta})+\sum_{i=1}^{t}\exp\left(H_{i|-i}\left(\bm{\beta}_{i} \mid 
		\bm{\beta}_{-i}\right)\right)\left(n_{i}+\gamma\right) \delta_{\bm{\beta}_{i}^{*}},\nonumber
	\end{equation}  
	with 
	\begin{equation}
		\begin{aligned}
			&V_{n_0}(t)=\sum_{k=1}^{\infty} \frac{k_{(t)}}{(\gamma k)^{(n_0)}} p_{K}(k); \quad \\
		&	H_{i|-i}\left(\bm{\beta}_{i} \mid 
			\bm{\beta}_{-i}\right)=\sum_{\{j:(i,j)\in E,j \neq i\}}  H_{ij} (\bm{\beta}_{i}
			\bm{\beta}_{j}),
		\end{aligned}
	\end{equation}
	where $\bm{\beta}^\ast_1,\ldots,\bm{\beta}^\ast_{t}$, $t\leq n_0-1$ are the distinct values taken by $\bm{\beta}_1,\ldots,\bm{\beta}_{n_0-1}$ and $n_i=\#\{j \in
	\{1,2,\ldots,n_0-1\}:\bm{\beta}_j= \bm{\beta}_{i}^\ast\}$, $x^{(m)}=x(x+1)
	\cdots(x+m-1)$ and $x_{(m)}=x(x-1) \cdots(x-m+1)$ .
\end{theorem}

	This theorem shows how the MRF constraints directly affect the  urn
	sampling scheme compared with MFM.  Considering the pairwise interactions, we
	model the conditional cost functions as
	\begin{equation}
		\label{cost}
		H_{i|-i}(\bm{\beta}_i \mid \bm{\beta}_{-i})=\lambda \sum_{\{j \in \partial (i)\}}
		I(\bm{\beta}_i=\bm{\beta}_j),
	\end{equation}
	where $\lambda$ is the smoothness parameter, $\partial (i)$ denotes the set of
	the neighbors of observation~$i$. The spatial smoothness can be controlled by
	the magnitude of $\lambda$. When $\lambda=0$, the MRF-MFM reduces to MFM
	\citep{miller2018mixture}. 
 The conditional cost function in \eqref{cost} is used in the data analysis of the paper.  

\subsection{Spatial Clustered Coefficient Regression for Count Value Data}
In the MRF-MFM, the multivariate normal distribution is a natural choice for the base distribution of $\bm{\beta}_1,\cdots,\bm{\beta}_k$. However, since the multivariate normal distribution is not a conjugate prior for Poisson regression, if it is used as the base distribution, it must be updated with Metropolis-Hastings or auxiliary parameters, such as those proposed by \citep{neal2000markov}, in Gibbs sampling algorithms. Additionally, the multivariate normal distribution has a thin tail, which is not suitable for estimating long-tail probabilities, such as the Poisson distribution. To address these limitations, \cite{bradley2018computationally} introduced the multivariate log-gamma distribution (MLG), which is conjugate with the Poisson distribution. The MLG distribution was derived from the multivariate gamma distribution formulated by \cite{demirhan2011multivariate} and transformed to the log-scale. However, this transformation complicates Gibbs sampling, as it requires component-wise updating to obtain known full-conditional distributions. Instead, the approach proposed by \cite{bradley2018computationally} allows for block-wise full-conditional distributions that are easier to simulate from. Furthermore, the MLG distribution exhibits asymptotic properties with respect to the multivariate normal distribution. A brief review of the MLG distribution is provided in the next section, and interested readers can refer to \cite{bradley2018computationally} for more details. In \cite{hu2018stat}, it is demonstrated that the MLG distribution has better long-tail probability properties compared to the multivariate normal distribution. In other words, the MLG distribution not only serves as a conjugate prior for the Poisson distribution but also has the ability to handle long-tailed probabilities. The ultimate goal is to propose an MRF-MFM for spatially clustered coefficients in Poisson regression based on the MLG prior.

\subsection{Probability Density Function for Multivariate Log-Gamma Distribution}
We first review the multivariate log-gamma distribution from \cite{bradley2018computationally}. We define the $n$-dimensional random vector $\bm{\phi}=(\phi_1,...,\phi_n)'$, which consists of $n$ mutually independent log-gamma random variables with shape and scale parameters organized into the $n$-dimensional vectors $\bm{\alpha} \equiv (\alpha_1,...,\alpha_n)'$, and $\bm{\kappa} \equiv (\kappa_1,...,\kappa_n)'$, respectively. Then define the $n$-dimensional random vector $\bm{q}$ as follows
\begin{equation}
	\bm{q}=\bm{\mu}+\bm{V}\bm{\phi},
	\label{linear transformation}
\end{equation}
where the matrix $\bm{V}\in \mathcal{R}^n\times\mathcal{R}^n$ and $\bm{\mu}\in\mathcal{R}^n $. \cite{bradley2018computationally} called $\bm{q}$ the multivariate log-gamma random vector. The random vector $\bm{q}$ has the following probability density function:
\begin{align}
\begin{aligned}
	f(\bm{q}\mid \bm{c},\bm{V},\bm{\alpha},\bm{\kappa})=\frac{1}{\text{det}(\bm{V})
	}\left(\prod_{i=1}^n \frac{\kappa_i^{\alpha_i}}{\Gamma(\alpha_i)
	}\right) \times\\
	\exp[\bm{\alpha}'\bm{V}^{-1}(\bm{q}-\bm{\mu})-\bm{\kappa}'\exp\{\bm{V}^
	{-1}(\bm{q}-\bm{\mu})\}]; \hspace{5pt}\bm{q}\in  \mathcal{R}^{n},
	\label{mlgpdf}
	\end{aligned}
\end{align}
where ``det'' represents the determinant function. As a shorthand we use the
notation, $\mathrm{MLG}\left(\bm{\mu}, \bm{V},
\bm{\alpha},\bm{\kappa}\right)$,  for the probability density function in
(\ref{mlgpdf}). 


\subsection{Conditional Distributions for Multivariate Log-Gamma Random Vectors}
Gibbs sampling from full-conditional distributions requires simulating from the conditional distributions of multivariate log-gamma random vectors. In this section, we provide a review of the technical results necessary for simulating from these conditional distributions.

We first look at Proposition 1 from \cite{bradley2018computationally}.
Let $\bm{q}\sim \mathrm{MLG}\left(\bm{c}, \bm{V}, \bm{\alpha},\bm{\kappa}\right)$, and let $\bm{q} = (\bm{q}'_1, \bm{q}'_2)'$, where $\bm{q}_1$ is $g$-dimensional and $\bm{q}_2$ is ($n-g$)-dimensional. In a similar manner, partition $\bm{V}^{-1}= \left[H\; B\right]$ into an $m\times g$ matrix H and an $m\times(m-g)$ matrix B. Then, the conditional pdf of $\bm{q}_1$ is given by
\begin{equation}
	f(\bm{q}_1\mid\bm{q}_2=\bm{d}, \bm{c}, \bm{\alpha},\bm{\kappa})= M
	\exp(\bm{\alpha}'\bm{H}\bm{q}_1-\bm{\kappa}_{1.2}'\exp(\bm{H}\bm{q}_1)).\label{MLGpdf}
\end{equation}
where $\bm{\kappa}_{1.2}\equiv \exp(\bm{B}\bm{d}-\bm{V}^{-1}\bm{c}-log(\bm{\kappa}))$ and the normalizeing constant M is
\begin{equation}
	M=\frac{1}{det(\bm{V}\bm{V}')^{\frac{1}{2}}}\left(\prod_{i=1}^{n}\frac{\kappa^{\alpha_i}_i}{\Gamma(\alpha_i)
	}\right)\frac{\exp{\bm{\alpha}'\bm{B}\bm{d}-\bm{\alpha}'\bm{V}^{-1}\bm{c}}}{\left[\int f(\bm{q}\mid \bm{c},\bm{V},\bm{\alpha},\bm{\kappa})d\bm{q}_1\right]_{\bm{q}_2=\bm{d}}},
\end{equation}
so the $\text{cMLG}(\bm{H},\bm{\alpha},\bm{\kappa}_{1.2})$ is equal to the pdf in equation
\eqref{cMLG}, where “cMLG” stands for “conditional multivariate log-gamma.” In \cite{bradley2018computationally}, it indicates that cMLG does not fall within the same class of pdfs given in \eqref{MLGpdf}. This is primarily due to the fact that the real-valued matrix H, within the expression of cMLG, is not square. Thus, we require an additional result that allows us to simulate from cMLG.

Next, we look at the Theorem~2 from \cite{bradley2018computationally}. Let
$\bm{q}\sim \mathrm{MLG}\left(\bm{0}_n, \bm{V}, \bm{\alpha},\bm{\kappa}\right)$, and
partition this $n$-dimensional random vector so that $\bm{q} = (\bm{q}'_1, \bm{q}'_2)'$,
where $\bm{q}_1$ is $g$-dimensional and $\bm{q}_2$ is ($n-g$)-dimensional. Additionally,
consider the class of MLG random vectors that satisfy the following:
\begin{equation}
	\bm{V}^{-1}=\left[Q_1\; Q_2\right] \begin{bmatrix} 
		R_1 & \bm{0}_{g,n-g} \\
		\bm{0}_{n-g,g} & \frac{1}{\sigma_2}\bm{I}_{n-g} 
	\end{bmatrix}
\end{equation}
where in general $\bm{0}_{r,t}$ is an $r\times t$ matrix of zeros; $\bm{I}_{n-g}$ is an
$(n-g)\times(n-g)$ identity matrix;
\begin{equation}
	\bm{H}=\left[Q_1\; Q_2\right] \begin{bmatrix} 
		R_1  \\
		\bm{0}_{n-g,g}  
	\end{bmatrix}
\end{equation}
is the QR decomposition of the $n\times g$ matrix H; the $n\times g$ matrix
$Q_1$ satisfies $Q'_1Q_1 = \bm{I}_g$, the $n\times(n-g)$ matrix $Q_2$ satisfies
$Q'_2Q_2 = \bm{I}_{n-g}$, and $Q'_2Q_1 = \bm{0}_{n-g,g}$; $R_1$ is a $g\times g$
upper triangular matrix; and $\sigma_2 > 0$. Hence, the marginal distribution
of the g-dimensional random vector $q_1$ is given by
\begin{equation}
	f(\bm{q}_1\mid\bm{H},\bm{\alpha},\bm{\kappa})= M_1
	\exp(\bm{\alpha}'\bm{H}\bm{q}_1-\bm{\kappa}'\exp(\bm{H}\bm{q}_1)).
	\label{cMLG}
\end{equation}
where the normalizing constant $M_1$ is
\begin{equation}
\begin{aligned}
	M_1=det(\left[\bm{H}\quad
	Q_2\right])\left(\prod_{i=1}^{n}\frac{\kappa^{\alpha_i}_i}{\Gamma(\alpha_i)
	}\right) \times\\
 \frac{1}{\int f(\bm{q}\mid \bm{0}_{n},\bm{V}=\left[\bm{H}\;	Q_2\right]^{-1},\bm{\alpha},\bm{\kappa})d\bm{q}_1}.
  \end{aligned}
\end{equation}
And, the $g$-dimensional random vector $\bm{q}_1$ is equal in distribution to $(\bm{H}'\bm{H})^{-1}\bm{H}'\bm{\omega}$, where the $n$-dimensional random vector $\bm{\omega} \sim \text{MLG}(\bm{0}_n, \bm{I}_n, \bm{\alpha}, \bm{\kappa})$. 

In \cite{bradley2018computationally}, it is evident that this particular class of marginal distributions (defined in Theorem 2 in \cite{bradley2018computationally}) falls into the same class of distributions as the conditional distribution of $\bm{q}_1$ given $\bm{q}_2$. And Theorem 2 in \cite{bradley2018computationally} provides a way to simulate from cMLG. Furthermore, it shows that it is (computationally) easy to simulate from cMLG provided that $g \ll n$. Recall that $\bm{H}$ is $n \times g$, which implies that computing the $g \times g$ matrix $(\bm{H}'\bm{H})^{-1}$ is computationally feasible when g is “small.” We refer the readers to \cite{bradley2018computationally} for a comprehensive discussion.



\section{Full Conditional Distributions and Algorithm}\label{fullcondit}
We adapt the MRF-MFM in conjunction with MLG to a spatial Poisson regression setting, focusing on the clustering of spatially-varying coefficients $\bm{\beta}(\bm{s}_1),\cdots, \bm{\beta}(\bm{s}_n)$, where
$\bm{\beta}(\bm{s}_i)$ is the $p$-dimensional coefficient vector for location $\bm{s}_i$. In our setting, we assume that the $n$ parameter vectors can be clustered into $k$ groups, i.e., $\bm{\beta}(\bm{s}_i) =\bm{\beta}_{z_i}\in\{\bm{\beta}_1,\cdots,\bm{\beta}_k\}$. Hence, the hierarchical model can be expressed as follows
\begin{align*}
	\begin{aligned}
	&	\textbf{Data Model: }	y(\bm{s}_i) \mid \bm{\beta}(\bm{s}_i) \sim
		\text{Poisson}(\exp\left(\bm{X}(\bm{s}_i)\bm{\beta}(\bm{s}_i))\right) \\
	&	\textbf{MRF: }	 (\bm{\beta}(\bm{s}_1),\cdots,\bm{\beta}(\bm{s}_n)) \sim
		M(\bm{\beta}(\bm{s}_1),\cdots,\bm{\beta}(\bm{s}_n))
		\prod_{i=1}^{n}G(\bm{\beta}(\bm{s}_i))\\
	&	\textbf{MLG: }		 \bm{\beta}_1,\ldots,\bm{\beta}_k \sim
		\text{MLG}(\bm{\mu},\bm{V},\bm{\alpha},\bm{\kappa})\\
	&	\textbf{MFM: }				 G(\bm{\beta}(\bm{s}_i))=\sum_{j=1}^k\pi_j\bm{\beta}_j,
		 \pi_1,\ldots,\pi_k \mid k \sim 
   \mbox{Dirichlet}(\gamma, \ldots, \gamma),\\
		& k \sim p(\cdot), \text{where $p(\cdot)$ is a p.m.f on } \{1,2, \ldots\}.
	\end{aligned}     
\end{align*}

The full conditional distributions in Markov chain Monte Carlo (MCMC) sampling
of MRF-MFM are given in Algorithm~\ref{Algorithm1} and Table~\ref{tab:para} where the detailed derivations are given in Section~\ref{sec:derivations} in the supplement.

\begin{algorithm}[ht]
	\caption{Collapsed sampler for MRF-MFM}\label{Algorithm1}
	\begin{algorithmic}
		\STATE {\bfseries Initialize:} $z=(z_1,\ldots,z_n)$ and $\bm{\beta}=(\bm{\beta}_1,\ldots,\bm{\beta}_k)$
		\FOR{$\text{each iteration}=1$ {\bfseries to} $B$}
		\STATE Update $\bm{\beta}=(\bm{\beta}_1,\ldots,\bm{\beta}_k)$ conditional on $z$ in a closed form as
		\begin{equation*}
			f(\pmb{\beta}_r\,|\, -)\sim\text{cMLG}(\bm{H}_{\beta},\bm{\alpha}_{\beta},\bm{\kappa}_{\beta})
		\end{equation*}
		where,
		
		$\bm{H}_{\bm{\beta}}=\begin{bmatrix}
			\bm{V^{-1}}\\
			\bm{X}(\bm{s}_i)
		\end{bmatrix}$
		$\bm{\alpha}_\beta=\begin{bmatrix}
			\alpha
			\\
			\sum_{z_i=r}^{}y(\bm{s}_i)\end{bmatrix}$
		$\bm{\kappa}_\beta=\begin{bmatrix}
			\kappa
			\\
			\sum_{z_i=r}^{}I_{(z_i=r)}\end{bmatrix}$
		\STATE Update $z=(z_1,\ldots,z_n)$ conditional on $\bm{\beta}=(\bm{\beta}_1,\ldots,\bm{\beta}_k)$ for each $i$ in (1,\ldots,n), we can get closed form expression for $P(z_i=c|z_{-i}, \bm{\beta})$:
		\begin{align*}
			\propto   
			\begin{cases}
				P(z_{i} = c \mid z_{-i}) d \text{Poisson}(y(\bm{s}_i),\exp(\bm{X}(\bm{s}_i)\bm{\beta}_c)) ,  \\
				\quad \quad \quad \text{at an existing table labeled}\, c\\.
				\frac{V_n(\left|{C_{-i}}+1)\right|}{V_n(\left|{C_{-i}}\right|)}\gamma m(y(\bm{s}_i)),   \quad \quad \, \text{if} \, c\,\text{is a new table}
			\end{cases}.
		\end{align*}
		where $C_{-i}$ denotes the partition obtained by removing $z_i$ and
		\begin{equation*}
			\begin{aligned}
				&m(y(\bm{s}_i))=\frac{1}{\mbox{det}(\bm{V}\bm{V}')^\frac{1}{2}}\left(\prod_{i=1}^{p}\frac{\kappa^{\alpha_i}_i}{\Gamma(\alpha_i)}\right)\frac{1}{M_1}
			\end{aligned}
		\end{equation*}
		where,
		\begin{equation*}
			\begin{aligned}
				&M_1=\mbox{det}(\left[\bm{H}_{\beta}\quad Q_2\right])\left(\prod_{i=1}^{n+p}\frac{\kappa^{\alpha_i}_i}{\Gamma(\alpha_i)}\right)\frac{1}{\int f(y(\bm{s}_i)\mid \bm{0},\bm{V}=\left[\bm{H}_{\beta}, Q_2\right]^{-1},\bm{\alpha},\bm{\kappa})}
			\end{aligned}
		\end{equation*}
		\ENDFOR
	\end{algorithmic}
\end{algorithm}

\begin{table}
	\center
	\caption{Parameters of the full conditional distribution}\label{tab:para}
	\begin{tabular}{cc}
		\hline
		Parameter&Form\\
		\hline
		$\bm{H}_\beta$& $\begin{bmatrix}
			\bm{V^{-1}}\\
			\bm{X}(\bm{s}_i)
		\end{bmatrix}$
		\\
		$\bm{\alpha}_\beta$& $\begin{bmatrix}
			\alpha
			\\
			\sum_{z_i=r}^{}y(\bm{s}_i)\end{bmatrix}$
		\\
		$\bm{\kappa}_\beta$&$\begin{bmatrix}
			\kappa
			\\
			\sum_{z_i=r}^{}I_{(z_i=r)}\end{bmatrix}$
		\\
		\hline
	\end{tabular}
	
\end{table} 
\subsection{Theoretical Properties under the Exchangeable Structure}

In this section,  we assume the covariates $\bm{X}(s_i)$ are generated from random homogenous distribution so it is marginalized. The incorporation of proper dependency structures into the estimation process and assessing uncertainty is always an interesting subject. However, complex dependency structures may destroy the consistency of MFM. Therefore, to maintain theoretical consistency, this paper considers the case in which samples from the base measure are \textit{a subset of an infinite sequence of exchangeable variables}.

In Bayesian Statistics, the infinite sequence of exchangeable random variables is an important
concept. When $\bm{\beta}_1,\ldots$ are infinite exchangeable,  for any finite
$k$,
\begin{equation}
	\bm{\beta}_1, \ldots, \bm{\beta}_k \stackrel{\mathcal{D}}{=} \bm{\beta}_{\pi(1)},
	\ldots, \bm{\beta}_{\pi(k)} \text { for all } \pi \in S(k),
\end{equation}
where $S(k)$ is the set of all permutations for the index set $\{1,\ldots,k\}$.
If $\bm{\beta}_1,\ldots$ are i.i.d. sampled from a distribution
$P(\bm{\beta})$, then they are exchangeable, but the reverse is not always true.
Some widely used models are based on  exchangeable random variables that are
not independent, like the P\'{o}lya's Urn \citep{blackwell1973ferguson} and
Gaussian random variables that have the same marginal distribution and the same
correlation between any two of them.

The famous de Finetti's Theorem \citep{de1929funzione} reveals the intrinsic
characterization of exchangeable random variables: there is a latent random
variable $\bm{\theta}$, such that $\bm{\beta}_1, \ldots, \bm{\beta}_n$  {are a subset of
	a infinite sequence of exchangeable variables} sampled from $\Pi(\bm{\beta}_1,
\ldots)$.
It is summarized into the
following sampling procedure:
\begin{equation}\label{eq:12}
	\bm{\theta} \sim \bm{\Theta}, \quad \bm{\beta}_1, \ldots, \bm{\beta}_k
	\stackrel{i.i.d.}{\sim} \Pi(\bm{\beta}|\bm{\theta}),
\end{equation}
where $\bm{\Theta}$ only depends on $\Pi(\bm{\beta}_1, \ldots)$. In other words, a
subset of an infinite sequence of exchangeable variables  are conditionally
i.i.d. given their latent labels. We refer to \cite{bernardo2009bayesian} for more details on exchangeable sequences.

\begin{theorem}\label{thm2}
	
	Suppose the data generating process follows equation~\eqref{MFM} with $H$ replaced by the hierarchical distribution in equation~\eqref{eq:12}, and the distribution is correctly specified. 	If $p_K(1),\ldots,p_K(k)>0$, denote $T$ as the random variable for the number of clusters and $t$ is all the possible values $T$ will take in the true data generating process. Then we have
	\begin{equation}
		\left|p\left(T=t \mid  \bm{y}\right)-p\left(K=k \mid \bm{y}\right)\right|
		\longrightarrow 0
	\end{equation}
	as $n \rightarrow \infty$.
	\label{thm:consistent}
\end{theorem}

Theorem~\ref{thm2} provide some insight into our proposed MRF-MFM, compared to
Dirichlet process mixture model with the above Markov random fields
\citep[DP-MRF;][]{orbanz2008nonparametric}. For DP-MRF, there could be a lot of
small spurious clusters due to inconsistency of the Dirichlet process mixture even in the i.i.d. case \citep{miller2013simple}. Due to the fact that we specify a prior distribution for the number of components, the number of components in the posterior is appropriately regularized. Even though the consistency result only holds for the exchangeable structure, we believe that the regularization effect holds for all types of structures. Theorem~\ref{thm2} is an extension of Theorem 5.2 in \cite{miller2018mixture} to the case of an exchangeable base measure. The limitation of the above theorem is that it does not explore the frequentist property of the posterior, where the number of clusters is assumed to be a fixed truth.

\section{Simulation}\label{sec:simu}

\subsection{Settings}\label{sec:bayes_comp}
Our goal is to sample from the posterior distribution of the unknown parameters
$k$, $z = (z_1,...,z_n) \in \{1,...,k\}$ and
$\bm{\beta}=(\bm{\beta}_1,\ldots,\bm{\beta}_k)$.  We choose $k-1 \sim
\mbox{Poisson}(1)$ and
$\gamma=1$, $\bm{\mu}=\bm{0}_n$, $\bm{V}=100\bm{I}_n$ and
$\bm{\alpha}=\bm{\kappa}=10000\bm{1}_n$ for all the simulations and real data
analysis, where $\bm{0}_n$ is an $n$-dimensional vector with 0, $\bm{1}_n$ is an
$n$-dimensional vector of~1's, and $\bm{I}_n$ is an $n$-dimensional identity matrix.
The computing algorithm and full conditional distributions are presented in Appendix \ref{fullcondit}, which efficiently
cycles through the full conditional distributions of $z_i | z_{-i}$ for $i =
1,2,\ldots,n$ and $\bm{\beta}$, where $z_{-i} = z\setminus{z_i}$. The marginalization over $k$ avoids the need for complicated reversible jump MCMC algorithms or allocation samplers. The posterior sampling algorithm is provided in Algorithm \ref{Algorithm1} in Appendix \ref{fullcondit}. Detailed deviations of the full conditional distributions are also outlined in Appendix \ref{fullcondit}. Using the posterior mean or median of clustering configurations $\bm{z}$ alone is not appropriate. Dahl's method \cite{dahl2006model} offers a remedy for posterior inference of clustering configurations based on the squared error loss. Additionally, alternative loss functions that do not rely on squared errors, such as those proposed in \cite{wade2018bayesian}, can be considered. The Rand Index (RI) \cite{rand1971objective} is used to measure the accuracy of clustering. The tuning parameter in the Markov random fields requires careful selection in our proposed model. We utilize the Logarithm of the Pseudo-Marginal Likelihood (LPML) \cite{ibrahim2013bayesian} for tuning parameter selection, where a model with a larger LPML value is preferred.
\subsection{Simulation Setting and Evaluation Metrics}\label{sec:simu_settings}
\begin{figure}[ht]
	\begin{center}
		\centerline{\includegraphics[width=\columnwidth]{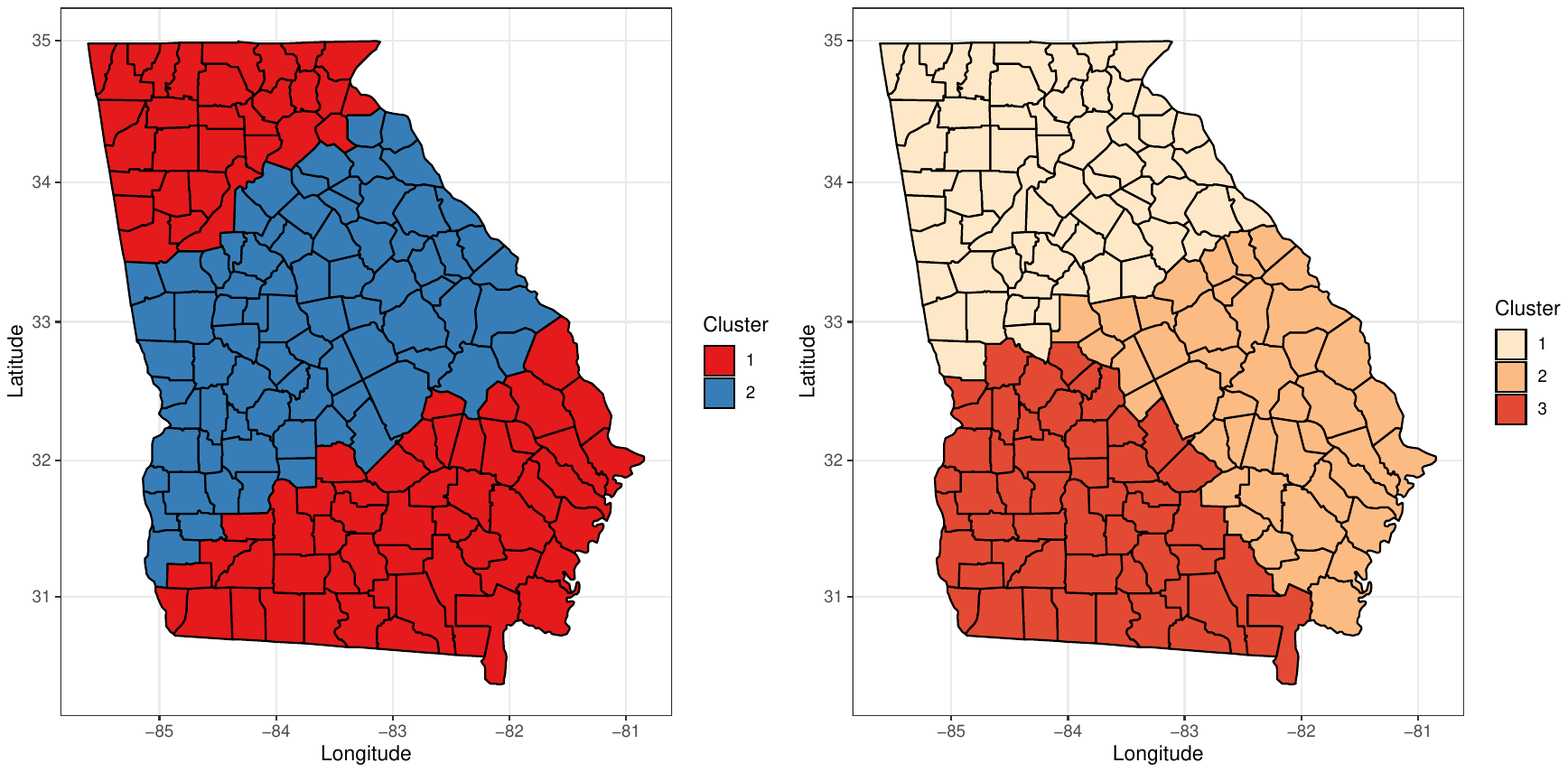}}
		\caption{Simulation design with two and three cluster assignments}	\label{fig:cluster:design}
	\end{center}
	\vskip -0.3in
\end{figure}

Our analysis is based on the spatial structure of the state of Georgia, which contains 159 counties. Using the county-level data, we build the graph using an adjacency matrix among different counties. 159 counties represent 159 vertices in this graph, and if a county shares a boundary with another county, then $v_i$ and $v_j$ are connected. This graph is used for both simulation studies and real data analysis. We consider two different spatial cluster designs shown in Figure~\ref{fig:cluster:design}. The first design consists of two disjoint parts located in the top and bottom parts of Georgia. A second cluster comprises the counties in the middle. The second design comprises three major spatial clusters. It is designed to mimic a common premature death pattern in which geographically distant areas can share a similar distribution pattern, and geographical proximity is not considered the only factor responsible for homogeneity in premature death rates.
%

Two different scenarios are considered for each design. The first scenario does not take into account spatial random effects, while in the second scenario, spatial random effects are included for each design. The spatial random effects are assumed to follow a multivariate normal distribution with a mean zero and exponential covariogram. Our simulation study consists of four scenarios in total. The details of the data generation process are given as
\begin{enumerate}
	\item $y(\bm{s}_i) \sim
	\text{Poisson}(X_1(\bm{s}_i)\beta_{1z_i}+X_2(\bm{s}_i)\beta_{2z_i})$, where
	$X_1(\bm{s}_i),X_2(\bm{s}_i) \stackrel{\text{ind}}\sim \text{Unif}(1,2)$,
	$i=1,\ldots,n$, $({\beta}_{11},{\beta}_{21})=(1,1)$,
	$({\beta}_{12},{\beta}_{22})=(1.5,1.5)$.
	
	\item $y(\bm{s}_i) \sim
	\text{Poisson}(X_1(\bm{s}_i)\beta_{1z_i}+X_2(\bm{s}_i)\beta_{2z_i}+w(\bm{s}_i))
	$, where $X_1(\bm{s}_i),X_2(\bm{s}_i) \stackrel{\text{ind}}\sim
	\text{Unif}(1,2)$, $i=1,\ldots,n$, $({\beta}_{11},{\beta}_{21})=(1,1)$,
	$({\beta}_{12},{\beta}_{22})=(1.5,1.5)$.
	$\omega\sim\text{N}(0,\sigma_{\omega}^{2}H(\phi))$, where
	$H(\phi)=\exp\left(-\phi\|s_i-s_j\|\right)$, we set $\sigma_{\omega}^{2}=0.3$
	and $\phi=0.05$.
	
	\item $y(\bm{s}_i) \sim
	\text{Poisson}(X_1(\bm{s}_i)\beta_{1z_i}+X_2(\bm{s}_i)\beta_{2z_i})$, where
	$X_1(\bm{s}_i),X_2(\bm{s}_i)\stackrel{\text{ind}} \sim \text{Unif}(1,2)$,
	$i=1,\ldots,n$, $({\beta}_{11},{\beta}_{21})=(0.5,0.5)$,
	$({\beta}_{12},{\beta}_{22})=(1,1)$, $({\beta}_{13},{\beta}_{23})=(1.5,1.5)$.
	
	\item   $y(\bm{s}_i) \sim
	\text{Poisson}(X_1(\bm{s}_i)\beta_{1z_i}+X_2(\bm{s}_i)\beta_{2z_i}+w(\bm{s}_i))
	$, where $X_1(\bm{s}_i),X_2(\bm{s}_i) \stackrel{\text{ind}}\sim
	\text{Unif}(1,2)$, $i=1,\ldots,n$, $({\beta}_{11},{\beta}_{21})=(0.5,0.5)$,
	$({\beta}_{12},{\beta}_{22})=(1,1)$, $({\beta}_{13},{\beta}_{23})=(1.5,1.5)$.
	$\omega\sim\text{N}(0,\sigma_{\omega}^{2}H(\phi))$, where
	$H(\phi)=\exp\left(-\phi\|s_i-s_j\|\right)$, we set $\sigma_{\omega}^{2}=0.3$
	and $\phi=0.05$.
\end{enumerate}
The  four scenarios are for two cluster design without spatial random effect, two cluster design with spatial random effect, three cluster design without spatial random effect, and three cluster design with spatial random effect, respectively.
In the three clusters design, the original regression coefficients are 
set to be 0.5, 1 and 1.5 for each cluster correspondingly. On the other hand, in
two clusters design, the original regression coefficient set to be 1 and 1.5
for each cluster, respectively. For each case, we add the spatial random
effect with the intensity. We use the centroid coordinate in each county to
represent that county then construct the spatial random effect. Also, the range
parameter and spatial variance parameter are both fixed in each simulation. In
each case, we avoid the zero count value to prevent numerical instability. Based on the estimated number of clusters and Rand Index (RI), the clustering performance is evaluated. Each replicate is also used to calculate the final number of clusters estimated. A total of 100 sets of data are generated under different scenarios. We run 5000 iterations of the MCMC chain and burn-in the first 1000 for each replicate.

\subsection{Simulation Results}\label{sec:simu_results} 
For each replicated data set, we fit MFM and MRF-MFM with different values of the smoothness parameter and select the best smoothness parameter for each replicate based on LPML. We see that our model outperforms the MFM model in terms of LPML in all four different scenarios. We also evaluate the performance in terms of estimation results of the number of clusters. We report the proportion of times the true cluster recovered among the 100 replicates. For the two-cluster without spatial random effect design, we find out our model can recover the true number of clusters 100\% of the replicates. And the MFM model can recover 85\% of the replicates. In this case, both models perform well in the number of clusters estimation. But our model outperforms the MFM model in terms of LPML value. For the two-cluster design with spatial random effects, we see that our model can recover the true number of clusters 97\% of the replicates, but the MFM model did not recover the true cluster for any replicates. For the three-cluster without spatial random effect design, we find out our model can recover the true number of clusters 88\% of the replicates. On the other hand, MFM recovers 62\% of the replicates. Finally, for the three-cluster design with spatial random effects, we find out our model can recover the true number of clusters 73\% of the replicates. However, MFM did not recover the true cluster for all replicates.

The results of the comparison of LPML, Rand index, and estimation of the number of clusters for each design can be found in Table~\ref{Tab:simuresults}. Our method can effectively estimate the true number of clusters based on the results shown in Table \ref{Tab:simuresults}. However, if spatial random effects exist, MFM will overestimate the number of clusters. Our proposed method also outperforms vanilla MFM with respect to model fitness and clustering, as demonstrated by the LPML values and Rand index.

\begin{table}[h]
	\caption{ Simulation Results for Four Scenarios including LPML, Rand Index (RI), and number of true cluster cover rate by MRF-MFM (optimal) model and MFM model. We provide mean and standard deviation for both LPML and RI.}
	\label{Tab:simuresults}
	\begin{tabular}{llccc}
		\toprule
		Scenario & Method  & LPML & RI & Cover Rate
		\\ \midrule
		\text{Scenario 1}  &\text{Optimal}&-544.29&0.9970&100\%\\
		&&(12.06)&(0.0062)&\\
		&\text{MFM}&-1146.32 &0.9901&85\%\\
		&&(593.33)&(0.0233)&\\ \midrule
		\text{Scenario 2}  &\text{Optimal}&-690.91&0.9875&97\%\\
				&&(34.36)&(0.0129)&\\
		&\text{MFM}&-7632.18&0.8348&0\%\\
				&&(1947.31)&(0.0597)&\\ \midrule
		\text{Scenario 3}  &\text{Optimal}&-752.76&0.9470&88\%\\
				&&(235.91)&0.0389&\\
		&\text{MFM}&-2201.69&0.9570&62\%\\
				&&(830.66)&(0.0231)&\\ \midrule
		\text{Scenario 4}  &\text{Optimal}&-1297.58&0.8469&73\%\\
				&&(278.33)&0.0434&\\
		&\text{MFM}&-8890.92&0.8350&0\%\\
				&&(2028.92)&(0.0431)&\\
		\bottomrule
	\end{tabular}
	\vskip -0.1in
\end{table}

Furthermore, we show the average mean square error (AMSE) of our proposed
method and MFM in Table \ref{Tab:AMSE}. We see that in all four different
scenarios, our proposed method outperforms MFM in terms of coefficients
estimations. The improvement of our proposed methods is evident for the data generated from the model with spatial random effect.

\begin{table}
	\caption{  AMSE for $\beta$ Estimation under All Scenarios}
	\label{Tab:AMSE}
	\begin{tabular}{llcc}
		\toprule
		 Method &  & \multicolumn{2}{c}{No Spatial Random effect}   \\ \cmidrule(lr){3-4}
		&&Two Clusters& Three Clusters
		\\ \midrule
		\text{MRF-MFM-MLG}  &$\hat{\beta_{1}}$&0.0848&0.2508\\
		&$\hat{\beta_{2}}$&0.0839&0.2435\\
		\text{MFM-MLG} &$\hat{\beta_{1}}$&0.1170&0.2841 \\
		&$\hat{\beta_{2}}$ &0.1164&0.2781\\
		\midrule
		&&\multicolumn{2}{c}{With Spatial Random effect} \\
		\cmidrule(lr){3-4}
		&&Two Clusters& Three Clusters\\
		\midrule
		\text{MRF-MFM-MLG} &$\hat{\beta_{1}}$&0.0966&0.3918\\
		&$\hat{\beta_{2}}$&0.0967&0.3814\\
		\text{MFM-MLG} &$\hat{\beta_{1}}$&0.3675&0.6996\\
		&$\hat{\beta_{2}}$&0.3668&0.6898\\
		\bottomrule
	\end{tabular}
	\vskip -0.1in
\end{table}


\section{Illustration: Premature Deaths in Georgia}\label{sec:real_data}
\subsection{Data Description}

\begin{figure}
	\centering
	\begin{subfigure}{0.47\textwidth}
		\includegraphics[width=\textwidth]{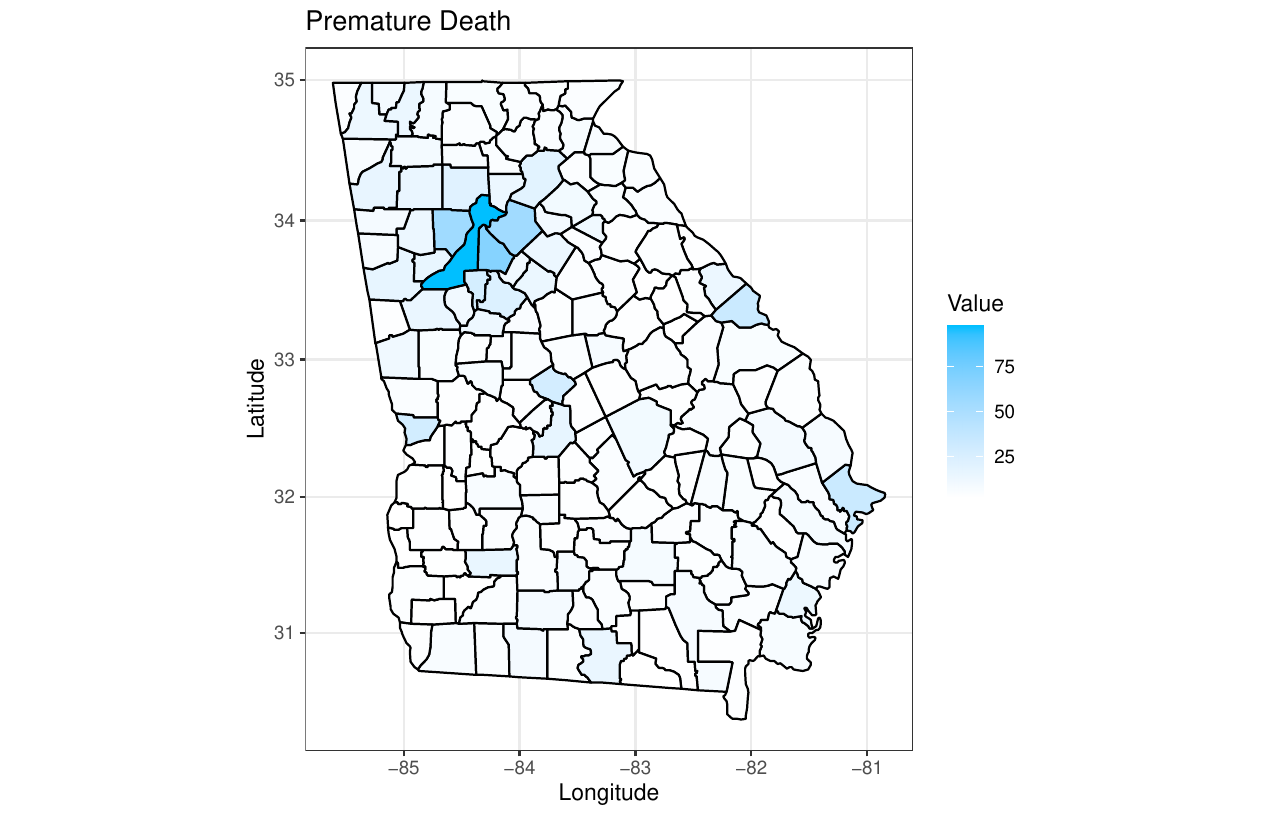}
		\caption{Count of the Premature Death (in hundreds)}
		\label{panel_response}
	\end{subfigure}
	\begin{subfigure}{0.47\textwidth}
		\includegraphics[width=\textwidth]{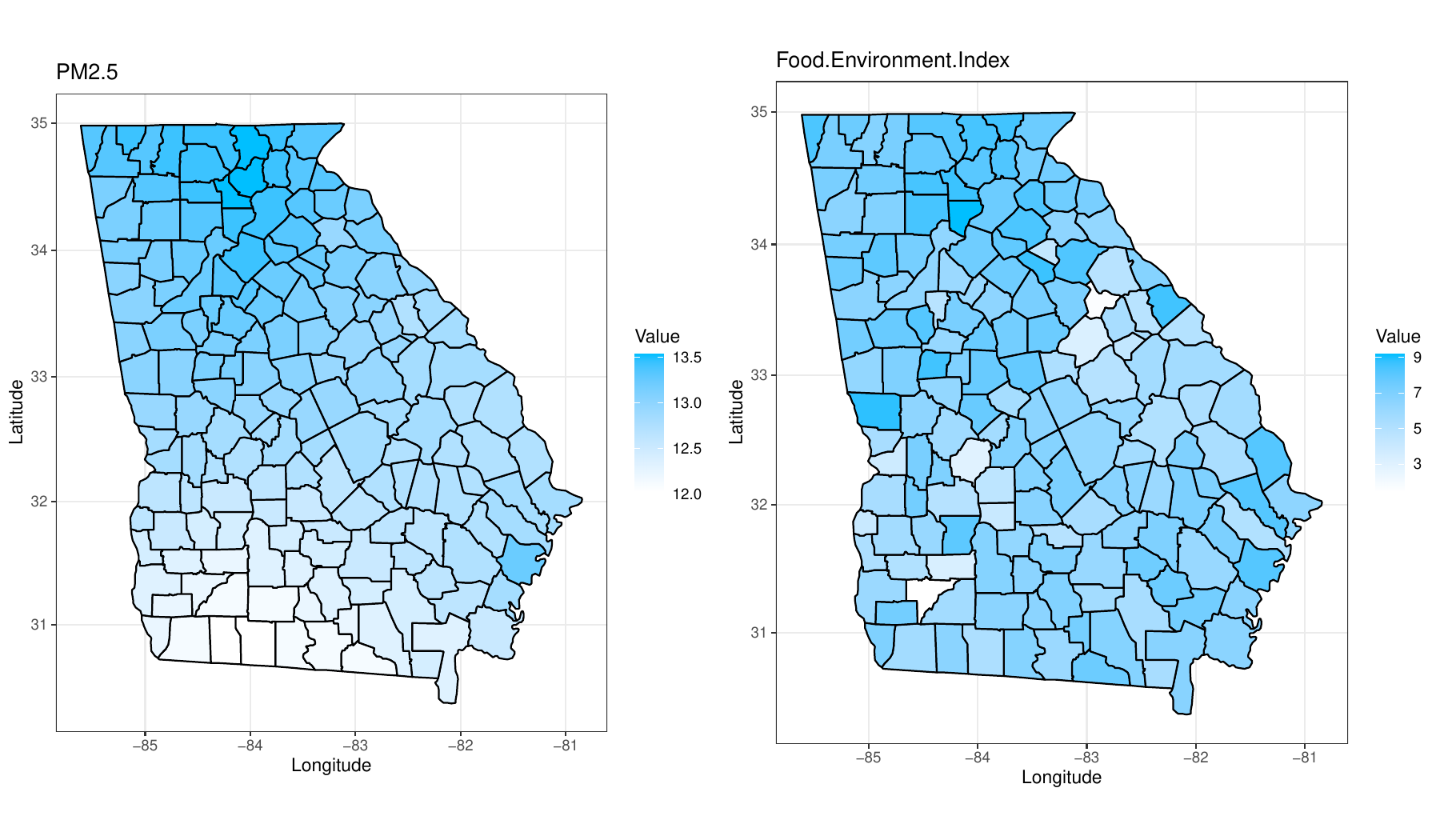}
		\caption{Visualizations of PM 2.5 and Food Environment Index}
		\label{panel_covariate}
	\end{subfigure}
	\label{fig:response_map}
\end{figure}

In this study, the proposed methods are used to analyze the factors that influence the number of premature deaths in Georgia. The objective of this study is to investigate the relationship between premature deaths and environmental factors such as PM 2.5 and food environment index. The dataset is available at \url{www.countyhealthrankings.org} with 159 observations corresponding to the 159 counties in state of Georgia in 2015. For each county, the dependent variable is the number of the premature death in each county. The premature death is the death that occurs before the
average age of death in a certain population. In the United States, the average age of
death is about 75 years. The dependent variable is the number of lives lost per
100,000 population before age 75 in each county. The two covariates we consider
in this paper are PM 2.5 ($X_1$) and food environment index ($X_2$). PM 2.5 is
the average daily density of fine particulate matter in micrograms per cubic
meter. The food environment index is the index of factors that contribute to a
healthy food environment, 0 (worst) to 10 (best). Figures \ref{panel_response} and
\ref{panel_covariate}  present a visualization of the response and two covariates on the
Georgia map.

\subsection{Data Analysis}
\begin{figure}[ht]
	\begin{center}
\includegraphics[width=\columnwidth]{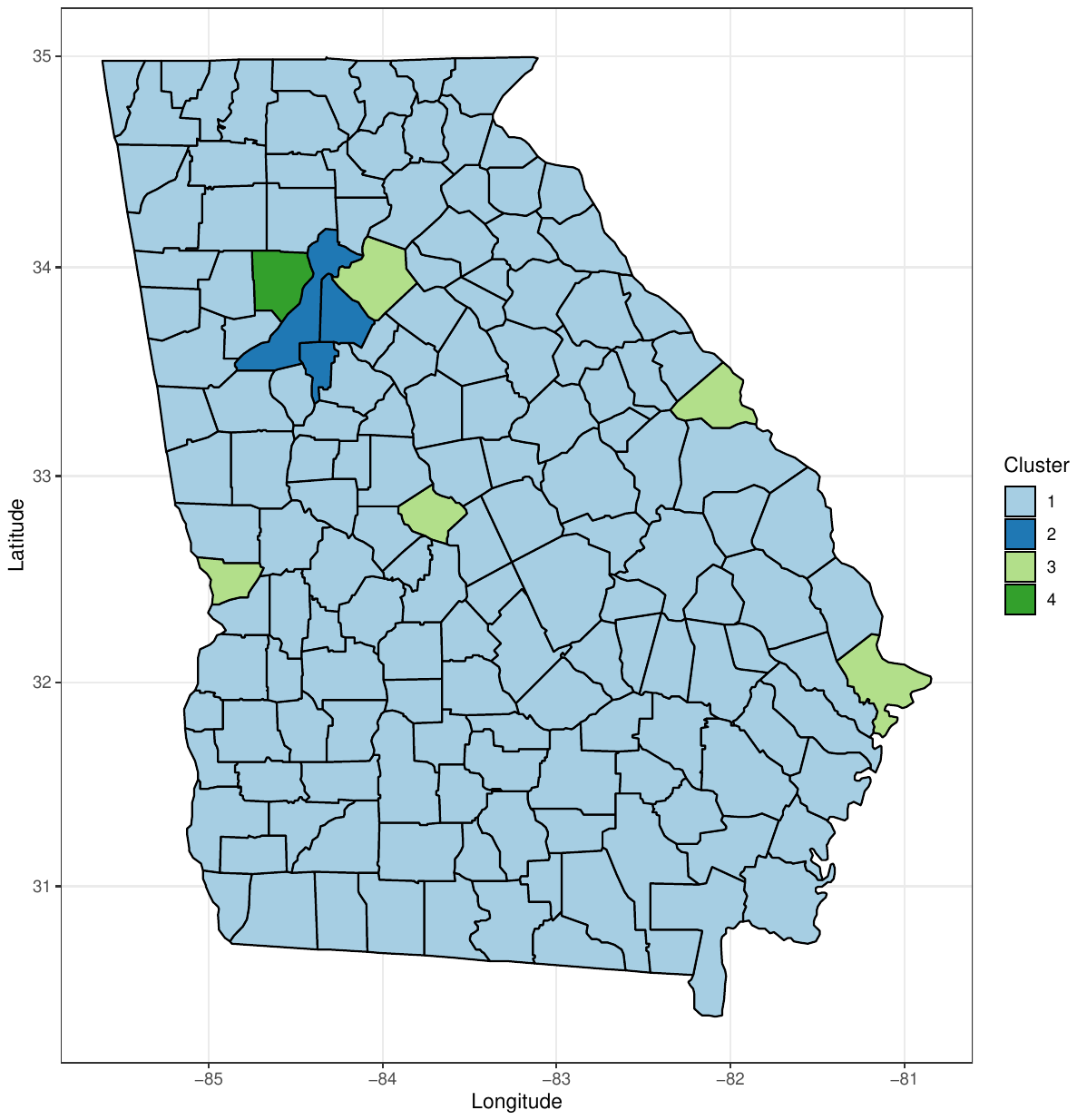}
			\includegraphics[width=\columnwidth]{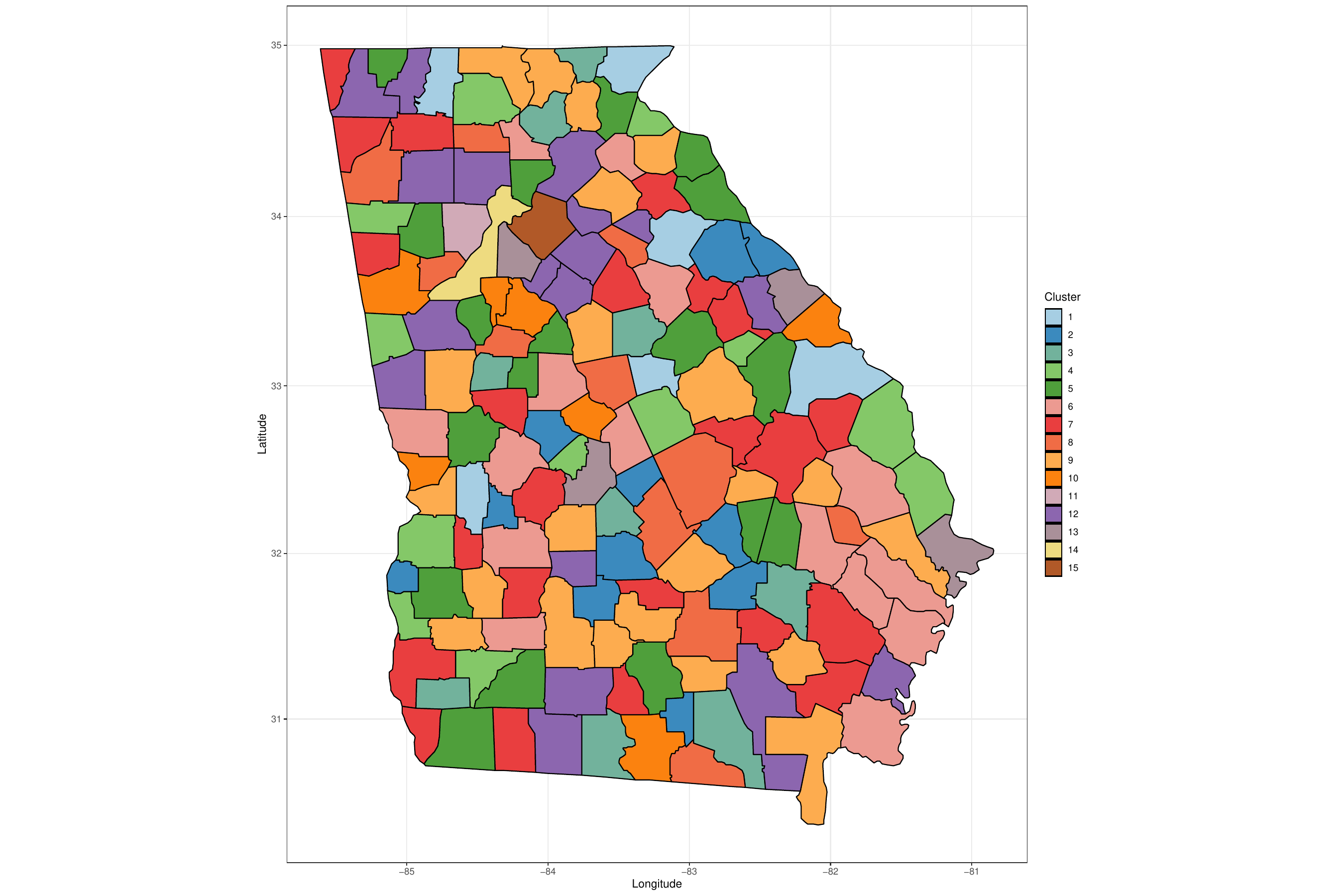}
		\caption{Top: Illustration of 4 clusters identified by the proposed method for counties. Bottom: Illustration of 15 clusters identified by vanilla MFM for counties.}	\label{fig:result}
	\end{center}

\end{figure}

In this section, we apply the proposed methodology to present a detailed
analysis of premature death data in the state of Georgia. First, we rescale the
data to a decent range as the variance in the Poisson distribution is
equal to the mean. The count of the premature death is scaled to hundreds.
We run 25,000 MCMC iterations and burn-in the first 15,000 iterations. The
smoothing parameter is tuned over the grid $\{0.1,0.2,\ldots,1\}$. All other
parameters are set to be consistent with the simulation study. The final
clustering result corresponds to the largest LPML
\citep{ibrahim2013bayesian}, hence we choose the smoothing parameter equal to 0.3.
The 159 counties turned out to be put into four clusters as illustrated in
Figure \ref{fig:result}. The number of the counties in each cluster are 150, 3,
5 and 1, respectively. We also compare our model with the best LPML to vanilla MFM, Latent
Gaussian Process (LGP) \citep{hadfield2010mcmc}, conditional autoregressive
(CAR) \citep{lee2013carbayes} models and Bayesian spatially varying coefficient models (SVC) \citep{gelfand2003spatial,wheeler2007assessment,finley2013spbayes}. The LPML values for candidate models are: -2221.45 (MRF-MFM), -3614.38 (MFM), -2461.31 (LGP), -5015.93 (CAR), -3123.47 (SVC). Based on the
LPML results, our proposed model outperforms other models. In contrast, there are 15 different clusters identified by vanilla MFM.  From
the estimation results shown in Table~\ref{Table:Estimation}, we see that all the
counties with higher PM 2.5 will have higher premature deaths. For Cobb County, PM 2.5 has the largest effect on premature death. {An extensive analysis could be conducted to investigate why the majority of counties are grouped into one cluster while the other three clusters only contain a couple of counties. For instance, one possible approach is to use log likelihood ratio test (LRT) to detect spatial cluster signals.}

\begin{table}

	\begin{center}
		\caption{ Dahl's method estimates for the four clusters of Georgia Data}
		\label{Table:Estimation}
		\begin{tabular}{cccc} 
			\toprule
			\textbf{Cluster} & $\hat{\beta_{0}}$& $\hat{\beta_{1}}$&$\hat{\beta_{2}}$\\
			\midrule
			\text{1}&-1.134  &0.077&  0.209  \\
			\text{2}&-3.644  &0.060 & 1.222  \\
			\text{3}&-1.325  &0.476 & -0.249  \\
			\text{4}&-0.188  &1.446 & -2.093 \\
			\bottomrule
		\end{tabular}
	\end{center}
\end{table}


\section{Discussions}\label{sec:discuss}
There are several topics beyond the scope of this paper that merit further investigation. Firstly, in our MCMC algorithm, a numerical integration is required for Gibbs sampling. Developing an efficient calculation algorithm for numerical integration would expand the applicability of our proposed methods. Additionally, the proposed algorithm encounters numerical instability when zero counts are observed, which should be addressed in future research. Moreover, different clusters may exhibit distinct sparsity patterns in the covariates. Incorporating spatially clustered sparsity structures of regression coefficients into the model would allow for the selection and identification of the most important covariates. The selection of a tuning parameter for the Markov random field is also necessary. Proposing a hierarchical model for the tuning parameter would be an interesting avenue for future work. Furthermore, exploring the frequentist properties of the posterior distribution is also an area that can be investigated in future research.

\section{Supplementary Materials}
The supplementary materials, including a detailed comparison with related literature such as \cite{li2019spatial} and \cite{anderson2017spatial}, proofs of the main theorems, derivations of full conditional distributions and MCMC algorithms, additional simulations using data from the states of Georgia and Mississippi, and reproducing codes for data analysis, are available on GitHub at \url{https://github.com/pengzhaostat/MLG_MFM}.

\begin{acks}  
Dr. Zhao's work is partially supported by NSF grant CCF-1934904. Dr. Hu's work is supported by NSF grants BCS-2152822, DMS-2210371, and SES-2243058.

\end{acks}

\bibliographystyle{ACM-Reference-Format}

\bibliography{sample}

\newpage
\appendix
\onecolumn

\section{Derivation of full conditionals}\label{sec:derivations}
For each term $\pmb{\beta}_r$ in
$\pmb{\beta}=(\pmb{\beta}_1,\ldots,\pmb{\beta}_k)$, the full conditional
distribution is:
\begin{equation}
	\begin{aligned}
		f(\pmb{\beta}_r\,|\, -) &\propto \text{MLG}(\bm{0}_p, \bm{V}, \bm{\alpha}, \bm{\kappa})
		\prod_{z_i=r}^{}\text{Poisson}(\exp\left(\bm{X}(\bm{s}_i)\bm{\beta}(s_{z_i}
		)\right)\\
		&\propto
		\exp(\bm{\alpha}'\bm{V}^{-1}\bm{\beta}_{r}-\bm{\kappa}'\exp(\bm{V}^{-1}\bm{\beta}_{r}))\prod_{z_i
			=r}^{}\exp\left(\bm{X}(\bm{s}_i)\bm{\beta}(s_{z_i})\right)^{y(\bm{s}_i)}\exp(
		-\exp\left(\bm{X}(\bm{s}_i)\bm{\beta}(s_{z_i})\right))\\
		&\propto
		\exp\left(\bm{\alpha}'\bm{V}^{-1}\bm{\beta}_{r}+\sum_{z_i=r}^{}y(\bm{s}_i)\bm{X}(\bm{s}
		_i)\bm{\beta}(s_{z_i})\right)\\
		&\exp\left(-\bm{\kappa}'\exp(\bm{V}^{-1}\bm{\beta}_{r})-\sum_{z_i=r}^{}I_{(z_i=r)}\exp
		(\bm{X}(\bm{s}_i)\bm{\beta}(s_{z_i}))\right)\\
		&\propto\exp\left[(\alpha,\sum_{z_i=r}^{}y(\bm{s}_i))'\begin{bmatrix}
			\bm{V^{-1}}\\
			\bm{X}(\bm{s}_i)
		\end{bmatrix}\beta_{r}
		\right]\exp\left[-(\kappa,\sum_{z_i=r}^{}I_{(z_i=r)})'\exp(\begin{bmatrix}
			\bm{V^{-1}}\\
			\bm{X}(\bm{s}_i)
		\end{bmatrix}\beta_{r})\right]
	\end{aligned}
\end{equation}

This implies that $f(\pmb{\beta}_r\,|\,
-)\sim\text{cMLG}(\bm{H}_{\beta},\bm{\alpha}_{\beta},\bm{\kappa}_{\beta})$.\\

For each term $z_i$ in $z=(z_i,\ldots,z_n)$, the full conditional distribution
is:
\begin{align*}
	P(z_{i} = c \mid z_{1}, \ldots, z_{i-1})  \propto   
	\begin{cases}
		P(z_{i} = c \mid
		z_{-i})d\text{Poisson}(y(\bm{s}_i),\exp(\bm{X}(\bm{s}_i)\bm{\beta}_r))  , &  \text{at
			table labeled}\, c\\
		\frac{V_n(|{C_{-i}}|+1)}{V_n(|{C_{-i}}|)}\gamma m(y(\bm{s}_i)),  &
		\, \text{if} \, c\,\text{is a new table}
	\end{cases}.
\end{align*}
where
\begin{equation*}
	\begin{aligned}
		m(y(\bm{s}_i))&=\int  \text{MLG}(\bm{0}_p, \bm{V}, \bm{\alpha},
		\bm{\kappa})\text{Poisson}(y(\bm{s}_i)\mid\bm{\beta}_r)d\bm{\beta}_r\\
		&\propto\int\frac{1}{det(\bm{V}\bm{V}')^\frac{1}{2}}\left(\prod_{i=1}^{p}\frac{\kappa
			^{\alpha_i}_i}{\Gamma(\alpha_i)}\right)\exp(\bm{\alpha}'\bm{V}^{-1}\bm{\beta}_{r}-\
		+\kappa'\exp(\bm{V}^{-1}\bm{\beta}_{r}))\\
		&\exp\left[\bm{X}(\bm{s}_i)\bm{\beta}_r\right]^{y(\bm{s}_i)}\exp\left[-\exp(\bm
		{X}(\bm{s}_i)\bm{\beta}_r)\right]\\
		&=\frac{1}{det(\bm{V}\bm{V}')^\frac{1}{2}}\left(\prod_{i=1}^{p}\frac{\kappa^{\alpha_i
			}_i}{\Gamma(\alpha_i)}\right)\\
		&\int\exp\left[(\alpha,\sum_{z_i=r}^{}y(\bm{s}_i))'\begin{bmatrix}
			\bm{V^{-1}}\\
			\bm{X}(\bm{s}_i)
		\end{bmatrix}\bm{\beta}_{r}
		\right]\exp\left[-(\kappa,\sum_{z_i=r}^{}I_{(z_i=r)})'\exp(\begin{bmatrix}
			\bm{V^{-1}}\\
			\bm{X}(\bm{s}_i)
		\end{bmatrix}\bm{\beta}_{r})\right]\\
		&=\frac{1}{det(\bm{V}\bm{V}')^\frac{1}{2}}\left(\prod_{i=1}^{p}\frac{\kappa^{\alpha_i
			}_i}{\Gamma(\alpha_i)}\right)\frac{1}{M_1}
	\end{aligned}
\end{equation*}
and
\begin{equation*}
	\begin{aligned}
		&M_1=det(\left[\bm{H}_{\beta}\quad
		Q_2\right])\left(\prod_{i=1}^{n+p}\frac{\kappa^{\alpha_i}_i}{\Gamma(\alpha_i)
		}\right)\frac{1}{\int f(y(\bm{s}_i)\mid \bm{0}_{n+p},\bm{V}=\left[\bm{H}_{\beta}\;
			Q_2\right]^{-1},\bm{\alpha},\bm{\kappa})}
	\end{aligned}
\end{equation*}
and $``\text{det}"$ is a short hand as determinant of a matrix.

\section{Additional Comparison for Simulation (State of Georgia)}
We present additional comparison for simulation section (State of Georgia). We compare our proposed method to LGP and CAR in two cluster design. In Figure 4, the values above zero indicate that our method has higher LPML than comparator. The results shown that we have a better result for both comparator.
\begin{figure}[ht]
	\begin{center}
		\centerline{\includegraphics[width=0.7\columnwidth]{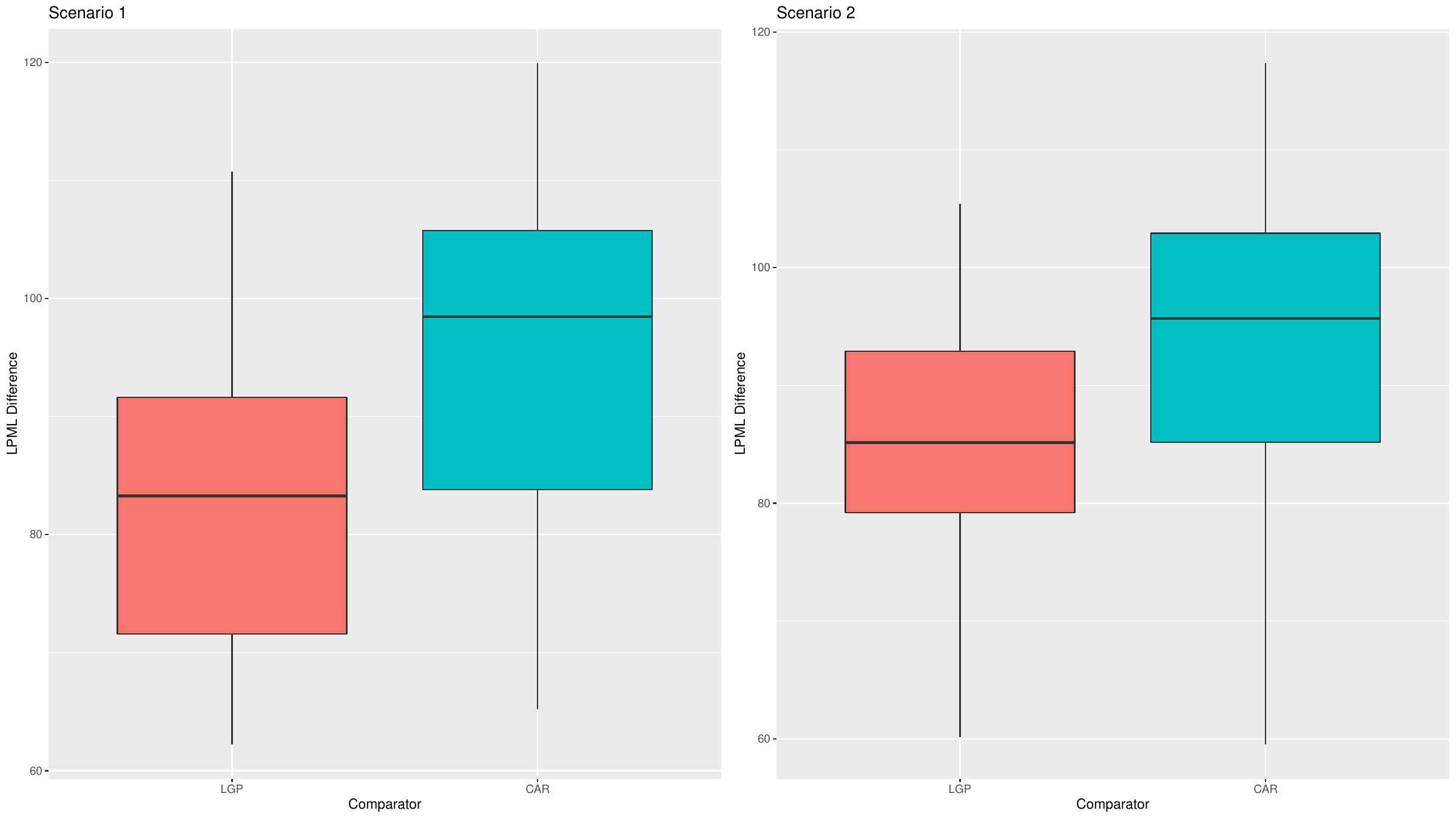}}
		\caption{Additional Comparison for Two Cluster Simulation (State of Georgia).}	
	\end{center}
	\vskip -0.3in
\end{figure}

\section{Additional Simulation for Different Spatial Graph (State of Mississippi)}
\begin{figure}[ht]
	\begin{center}
		\centerline{\includegraphics[width=0.7\columnwidth]{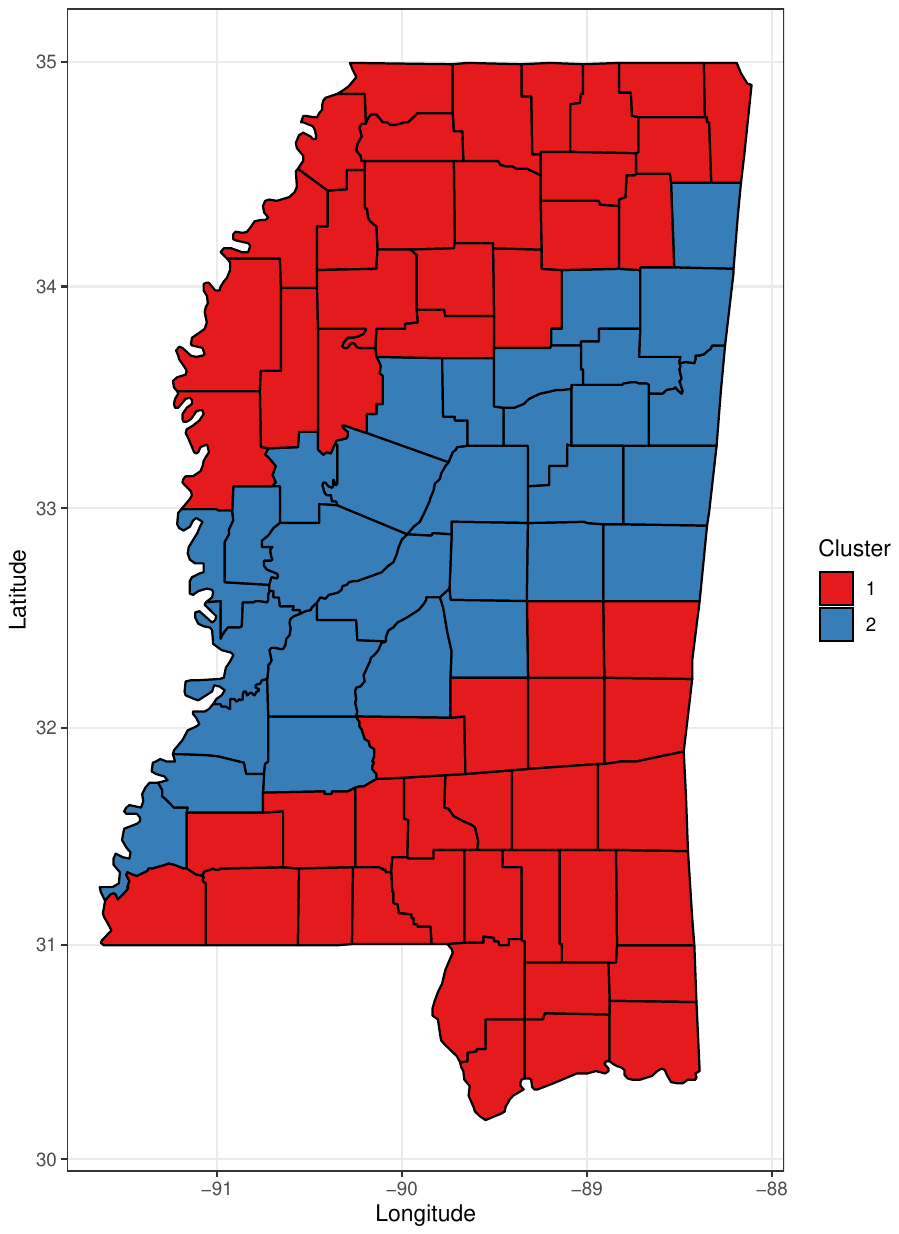}}
		\caption{Simulation design with two cluster assignments. (State of Mississippi)}	\label{fig:cluster:designMS}
	\end{center}
	\vskip -0.3in
\end{figure}

We provide another simulation design with different spatial graph. This additional analysis is based on the spatial structure of the state of Mississippi, which contains 82 counties. We consider a different spatial cluster designs shown in Figure~\ref{fig:cluster:designMS}. This design consists of two disjoint parts located in the top and bottom parts of Mississippi. 

Two different scenarios are considered. The first scenario does not take into account spatial random effects, while in the second scenario, spatial random effects are included for each design. The spatial random effects are assumed to follow a multivariate normal distribution with a mean zero and exponential covariogram. Based on the estimated number of clusters and Rand Index (RI), the clustering performance is evaluated. Each replicate is also used to calculate the final number of clusters estimated. A total of 50 sets of data are generated under different scenarios. We run 3000 iterations of the MCMC chain and burn-in the first 1000 for each replicate.

The results of the comparison of LPML, Rand index, and estimation of the number of clusters for each design can be found in Table~\ref{Tab:simuresultsMS}. Our proposed method outperforms vanilla MFM with respect to model fitness and clustering, as demonstrated by the LPML values and Rand index. Additional comparison to LGP and CAR also presented. In Figure 6, the values above zero indicate that our method has higher LPML than comparator. The results shown that we have a better result for both comparator. 

\begin{table}[ht]
	\caption{Simulation Results including LPML, Rand Index (RI), and number of true cluster cover rate (CR) by MRF-MFM (optimal) model and MFM model. We provide mean and standard deviation for both LPML and RI.}
	\label{Tab:simuresultsMS}
	\begin{tabular}{lcccccccc}
		\toprule
	 Method &	Scenario  & LPML & RI & CR & Scenario & LPML & RI & CR
		\\ \midrule
	\text{Optimal} &	\text{1}  &-295.79&0.9954&100\% &	\text{2} &-291.76&0.9966&100\%\\
		&&(9.29)&(0.0179)& &&(10.93)&(0.0135)&\\
	\text{MFM}	&&-819.39 &0.9901&98\%  & &-727.47&0.9901&96\%\\ 
		&&(407.15)&(0.0257)&   &&(272.14)&(0.0269)&\\
		\bottomrule
	\end{tabular}
\end{table}

\begin{figure}[ht]
	\begin{center}
		\centerline{\includegraphics[width=0.7\columnwidth]{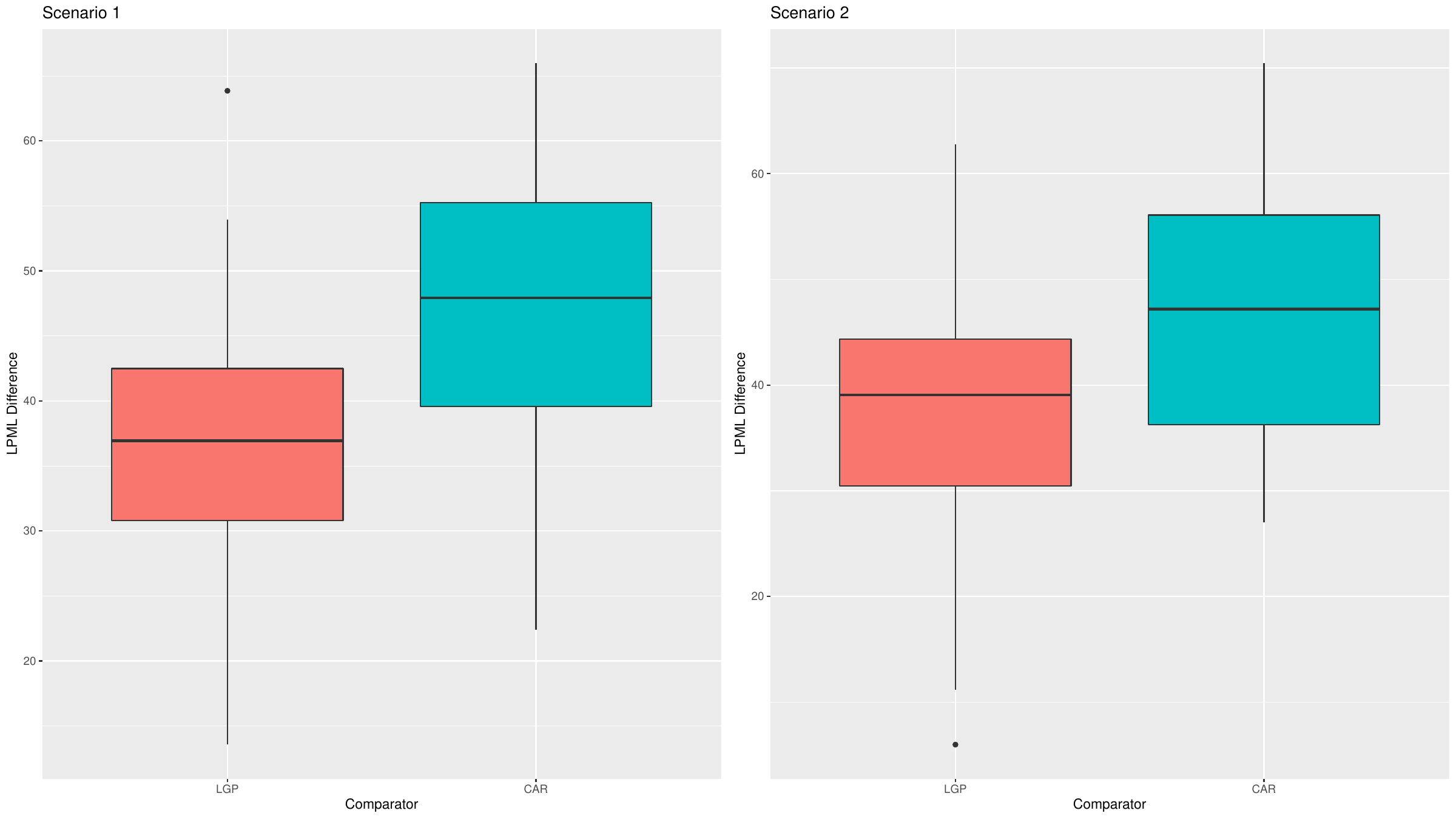}}
		\caption{Additional Comparison for Two Cluster Simulation (State of Mississippi).}	
	\end{center}
	\vskip -0.3in
\end{figure}

\section{Comparison with the Related Literature}\label{sec:comparison}

Both \cite{li2019spatial} and \cite{anderson2017spatial} perform clustered coefficients separately for different coefficients, as shown in Figures 2 and 4 in \cite{li2019spatial}, and Figures 1 and 3 in \cite{anderson2017spatial}. In contrast, our approach performs clustering for all coefficients together, where all coefficients share the same cluster configuration. For example, Figure~\ref{fig:result} illustrates our approach generating a single clustered configuration on the county map, despite estimating three coefficients. It is important to correctly specify whether a single partition or different partitions should be used for a model in practice. Here, we consider a single partition for different coefficients for the following reasons:

\begin{itemize}
    \item Methodology: Using a single partition for different coefficients is more interpretable as it automatically produces a cluster assignment of the spatial objects.
    \item Simulation: When the true data generation process is based on a single cluster assignment, a single partition with different coefficients performs better than having separate partitions. This is because the latter leads to more variance in estimations, resulting in overfitting the model. To demonstrate this, we implemented \cite{li2019spatial}'s approach using the R package \textit{genlasso} and compared the average mean squared errors (AMSE) for a two-cluster design without spatial random effect. Our estimation error is around 0.085 according to Table 2, while the estimation error of \cite{li2019spatial}'s approach is 26.48. The reproducible code, "fusedlasso.R," is provided in the supplement. The substantial difference in estimation errors is due to (1) our model being correctly specified with the Poisson likelihood, while \textit{genlasso} uses a Gaussian likelihood, and (2) our model being correctly specified with the single cluster assignment.
    \item Theoretical details: Single clusters for different coefficients are less common in the literature, possibly because they are mathematically more complex. For example, in penalized regression (\cite{li2019spatial}), a simple penalty on the aggregated coefficients is sufficient to induce separate clusters, corresponding to the fused lasso. However, inducing a common cluster assignment among different coefficients requires considering group-wise penalizations, corresponding to a more complex type of regularization.
    \item Application: For our analysis of Georgia's premature death data, the coefficients of interest, such as PM 2.5 and the food environment index, are primarily influenced by environmental factors. Therefore, it makes sense to claim that their effects cluster in homogeneous groups.
\end{itemize}

\section{Proof of the Theorem~\ref{thm:urn}}\label{sec:proof1}

By Bayes' theorem, we have:
\begin{equation}
	\Pi\left(\bm{\beta}_{i} | \bm{\beta}_{-i}\right) \propto \Pi\left(\bm{\beta}_{1},
	\ldots, \bm{\beta}_{n_0}\right) = P\left(\bm{\beta}_{1}, \ldots, \bm{\beta}_{n_0}\right)
	M\left(\bm{\beta}_{1}, \ldots, \bm{\beta}_{n_0}\right) \propto  P\left(\bm{\beta}_{i}|
	\bm{\beta}_{-i}\right) M\left(\bm{\beta}_{i} |\bm{\beta}_{-i}\right).
\end{equation}
As shown in \cite{miller2018mixture}, by conditioning on the different
possible situations of the cluster for the new observations, we have
\begin{equation}
	P\left(\bm{\beta}_{i}| \bm{\beta}_{-i}\right) \propto \frac{V_{n_0}(t+1)
		\gamma}{V_{n_0}(t)}  P(\bm{\beta}_i)+\sum_{i=1}^{t} \left(n_{i}+\gamma\right)
	\delta_{\bm{\beta}_{i}^{*}}.
\end{equation}
Let $\partial(i):=\{j:(i,j)\in E\}$. When considering the full conditional
distribution
\begin{equation}
	M\left(\bm{\beta}_{i}|\bm{\beta}_{-i}\right) \propto \exp \left(H_{i|-i}(\bm{\beta}_i |
	\bm{\beta}_{-i})\right),
\end{equation}
where $H_{i|-i}(\bm{\beta}_i | \bm{\beta}_{-i})$ only depends on $H_{ij}(\bm{\beta}_i
\bm{\beta}_j)$ for $(i,j)\in E$.
Note that 
\begin{equation}\label{H:characteristic}
	\begin{array}{ll}{H_{i|-i}\left(\bm{\beta}_{i} | \bm{\beta}_{-i}\right)=0} & {\text {
				if } S_{i} \notin S_{\partial(i)}}\end{array},
\end{equation}
where $\bm{s}_i$ specifies the cluster that $\bm{\beta}_i$ belongs to.
With the property in equation~\eqref{H:characteristic} and the assumption that
$P$ is continuous, $\exp \left(H_{i|-i}\left(\bm{\beta}_{i} |
\bm{\beta}_{-i}\right)\right)=1$ almost surely for $\bm{\beta}_{i} \sim P$. Then
given any measurable function $f$ for $P(\bm{\beta})$ and any subset $A$ for the domain of
$\bm{\beta}_i$,
\begin{equation}
	\begin{aligned}{\int_{A} f\left(\bm{\beta}_{i}\right) M\left(\bm{\beta}_{i} |
			\bm{\beta}_{-i}\right) P\left(\bm{\beta}_{i}\right) d \bm{\beta}_{i}}  {=\int_{A}
			f\left(\bm{\beta}_{i}\right) \frac{1}{Z_{H'}} \exp \left(H_{i|-i}\left(\bm{\beta}_{i}
			| \bm{\beta}_{-i}\right)\right) P\left(\bm{\beta}_{i}\right) d \bm{\beta}_{i}} \\
		{\quad=\int_{A} f\left(\bm{\beta}_{i}\right) \frac{1}{Z_{H'}}
			P\left(\bm{\beta}_{i}\right) d \bm{\beta}_{i}} ,\end{aligned}
\end{equation}
where the constant $Z_{H'}$ only depends on the constant part of
$M\left(\bm{\beta}_{i} | \bm{\beta}_{-i}\right)$.
Hence, the full conditional of $\Pi$ can be derived
\begin{equation}
	\Pi\left(\bm{\beta}_{i} | \bm{\beta}_{-i}\right) \propto \frac{V_{n_0}(t+1)
		\gamma}{V_{n_0}(t) } P(\bm{\beta}_i)+\sum_{i=1}^{t} \exp\left(H_{i|-i}\left(\bm{\beta}_{i} |
	\bm{\beta}_{-i}\right)\right)\left(n_{i}+\gamma\right) \delta_{\bm{\beta}_{i}^{*}}.
\end{equation}       

   \section{Proof of the Theorem~\ref{thm:consistent}}\label{sec:proof3}

   \begin{proposition}\label{prop1}
	If the data generating process follows equation~\eqref{MFM} with $H$ replaced by the hierarchical distribution in equation~\eqref{eq:12}, then we have
	\begin{align}
		p(\mathcal{C}) &= V_n(t) \prod_{ c\in \mathcal{C}} {\gamma^{(|c|)}}, \quad
		p(\mathcal{C} \mid k) = \frac{k_{(t)}}{(\gamma k)^{(n)}} \prod_{ c\in \mathcal{C}}  {\gamma^{(|c|)}},\nonumber\\
		\quad p(K=t \mid T=t) &= \frac{t_{(t)}}{V_n(t)(rt)^{(n)}}p_K(t) \rightarrow 1, \quad
		\mathcal{C} \perp K \mid T,
	\end{align}
	where $t=|\mathcal{C}|$ is the number of clusters while $T$ is the corresponding random variable of $t$ and $V_n(t)$ is defined in equation~\eqref{vt}.
\end{proposition}
The proof of this proposition directly follows from~\cite{miller2018mixture}, since all conclusions only involves on $\mathcal{C},K$ and $T$, while the i.i.d assumption on $\bm{\beta}$ is not used. 
 
 \begin{lemma}\label{thm:indpendent} Suppose the data generating process in Proposition~\ref{prop1}, such that the distribution is correctly specified.  Given the cluster
	configuration $\mathcal{C}$, the data~$\bm{y}$ and the number of components $K$ are
	independent.
\end{lemma}
\begin{remark}
As with MFM, we generalize the same result to exchangeable cases. Since the dependence between $\bm{y}$ is totally decided by $\bm{\beta}$, when $\bm{\beta}$ are exchangeable, all the $\bm{\beta}$ play the same role in generating $\bm{y}$. When $\bm{\beta}$ are marginalized, the cluster configuration $\mathcal{C}$ covers the same information with the number of components $K$ and the latent labels $\bm{z}$.
\end{remark}

\subsection{Proof of the Lemma~\ref{thm:indpendent}}\label{sec:proof2}
\begin{proof}
We show that  the conditional independence among data $\bm{y}$ and the number of
components $K=k$ given the cluster configuration $\mathcal{C}$ still holds when
all $\bm{\beta}$ are exchangeable. \\
Let $E_i=\{j:z_j=i\}$, based on the definition of $E_i$ and $\bm{z}$, we have
\begin{equation}
	p(\bm{y}|\bm{\beta},\bm{z},k)= \prod_{i=1}^{k} \prod_{j \in E_i} p(y_j |\bm{\beta}_i) =
	\prod_{i=1}^{t} \prod_{j \in E_i} p(y_j |\bm{\beta}_i^\ast),
\end{equation}   where $\bm{\beta}_i^*$, $i=1,2,\ldots,t$ are the distinct values
of $\bm{\beta}_{1:k}$ decided by $\bm{z}$ and $\bm{y}$, and
$\bm{\beta}_{1:k}=(\bm{\beta}_1,...,\bm{\beta}_k)^{\top}$. Given $\bm{z}$, the
transformation from variable $\bm{\beta}_{1:k}$ to $\bm{\beta}_{1:t}^\ast$, is
totally decided, so when marginalizing the unused $\bm{\beta}_{(t+1):k}^\ast$,
given any function $g(\bm{\beta}_{1:t}^\ast)$, we have the identity
\begin{equation}
	\int_{\Theta^k} g(\bm{\beta}_{1:t}^\ast) p(\bm{\beta}|\bm{z},k) \left(d \bm{\beta}\right) =
	\int_{\Theta^t} g(\bm{\beta}_{1:t}^\ast) p(\bm{\beta}^\ast_{1:t})  d \bm{\beta}^\ast.
\end{equation}
Note that $\bm{\beta}^*_{1:t}$ are exchangeable based on assumption, then the
density after marginalizing $\bm{\beta}$ can be seen
\begin{equation}
	\begin{aligned} p(\bm{y} | \bm{z}, k) &=\int_{\Theta^{k}} p(\bm{y} | \bm{\beta}, \bm{z}, k) p(
		\bm{\beta} |\bm{z}, k) d\bm{\beta}= \int_{\Theta^k}\prod_{i=1}^{k} \prod_{j \in E_i}
		p(y_j |\bm{\beta}_i)p(\bm{\beta}|\bm{z},k) \left(d \bm{\beta}\right) \\
		&= \int_{\Theta^t}\prod_{i=1}^{t} \prod_{j \in E_i} p(y_j
		|\bm{\beta}^\ast_i)p(\bm{\beta}^\ast_{1:t})  d \bm{\beta}^\ast \\
		& \stackrel{(i)}{=} \int_{\Theta^t}\prod_{i=1}^{t}  p(\bm{y}_{E_i}
		|\bm{\beta}^\ast_i) \int \prod_{i=1}^t p(\bm{\beta}_i^*|\bm{\theta}) d F(\bm{\theta})  d
		\bm{\beta}^\ast  \\
		& \stackrel{(ii)}{=} \int \int_{\Theta^t}\prod_{i=1}^{t} \left[ p(\bm{y}_{E_i}
		|\bm{\beta}^\ast_i)  p(\bm{\beta}_i^*|\bm{\theta}) \right] d \bm{\beta}^\ast  d F(\bm{\theta})
		\\
		& \stackrel{(iii)}{=} \int \prod_{i=1}^{t} m_i(\bm{y}_{E_i},\bm{\theta})  d
		F(\bm{\theta}), \end{aligned}
\end{equation}
where $m_i(\bm{y}_{E_i},\bm{\theta})$ is a function only depends on $\bm{y}_{E_i}$ and
$\bm{\theta}$. In addition, $(i)$ directly follows from de Finetti's Theorem; for
$(ii)$, we apply the Fubini's theorem;  $(iii)$ is because the expression
depends only on $\bm{z},k$ through $\mathcal{C}=\mathcal{C}(\bm{z})$ since there is
no correspondence between $E_i$ and $\bm{\beta}_i^*$ after integrating out
$\bm{\beta}^*$. From the last expression, we can see $p(\bm{y} | \bm{z}, k) $ can be
represented as a function of $\mathcal{C},\bm{y}$, which implies that $\bm{y}$ and
$K$ are conditional independent given the cluster configuration $\mathcal{C}$.
\end{proof}

Based on the fourth line in Proposition 1 and Lemma~\ref{thm:indpendent}, we
have $\mathcal{C} \perp K \,|\, T $ and $\bm{y} \perp  K \,|\,\mathcal{\bm{C}}$. Then
we have
\begin{equation}
	p(\bm{y} | t, k)=\sum_{\mathcal{C} :|\mathcal{C}|=t} p(\bm{y} | \mathcal{C}, t, k)
	p(\mathcal{C} | t, k)=\sum_{\mathcal{C} :|\mathcal{C}|=t} p(\bm{y} | \mathcal{C},
	t) p(\mathcal{C} | t)=p(\bm{y} | t),
\end{equation}
which implies $\bm{y} \perp K \,|\, T$. Then for any $n \geq k$, 
\begin{equation}
	p\left(K=k \,|\, \bm{y} \right)=\sum_{t=1}^{k} p\left(K=k \,|\, T=t, \bm{y}\right)
	p\left(T=t \,|\, \bm{y}\right)=\sum_{t=1}^{k} p\left(K=k \,|\, T=t\right)
	p\left(T=t \,|\, \bm{y}\right).
\end{equation}
In addition, $p(K=t | T=t)=1/{V_{n}(t)} \longrightarrow 1$ as $n \rightarrow
\infty$ based on the third equation in Proposition 1. Thus
\begin{equation}
	p\left(K=k \,|\, \bm{y} \right) \rightarrow \sum_{t=1}^{k} I(k=t)p(T=t \,|\,\bm{y} )
	= p(T=t\,|\,\bm{y}).
\end{equation}



\end{document}